\documentclass[prb,aps,epsf,twocolumn,showpacs,10pt,superscriptaddress ]{revtex4-2}
\usepackage{graphicx,amsfonts,times,bm,amsmath,verbatim,color,array}
\newcommand{\<}{\langle}
\newcommand{\e}{\varepsilon}
\renewcommand{\>}{\rangle}

\usepackage{natbib}
\usepackage{graphicx,amssymb,amsmath,ifthen,braket,xcolor}
\usepackage[linktocpage=true,colorlinks=true,urlcolor=blue,linkcolor=red,citecolor=blue]{hyperref}
\usepackage{bm}

\newcommand{\angstrom}{\mbox{\normalfont\AA}}
\usepackage{xr}
\makeatletter
\newcommand*{\addFileDependency}[1]{
  \typeout{(#1)}
  \@addtofilelist{#1}
  \IfFileExists{#1}{}{\typeout{No file #1.}}
}
\makeatother

\newcommand*{\myexternaldocument}[1]{
    \externaldocument{#1}
    \addFileDependency{#1.tex}
    \addFileDependency{#1.aux}
}
\myexternaldocument{SM}

\begin{document}

\title{Band tilt-induced nonlinear Nernst effect in topological insulators: An efficient generation of high-performance spin polarization}
\date{\today}

\author{Chuanchang Zeng}
\affiliation{Centre for Quantum Physics, Key Laboratory of Advanced Optoelectronic Quantum Architecture and Measurement(MOE),
School of Physics, Beijing Institute of Technology, Beijing, 100081, China}
\affiliation{Beijing Key Lab of Nanophotonics $\&$ Ultrafine Optoelectronic Systems,
School of Physics, Beijing Institute of Technology, Beijing, 100081, China}

\author{Xiao-Qin Yu}
\affiliation{School of Physics and Electronics, Hunan University, Changsha 410082, China}
\author{Zhi-Ming Yu}
\affiliation{Centre for Quantum Physics, Key Laboratory of Advanced Optoelectronic Quantum Architecture and Measurement(MOE),
School of Physics, Beijing Institute of Technology, Beijing, 100081, China}
\affiliation{Beijing Key Lab of Nanophotonics $\&$ Ultrafine Optoelectronic Systems,
School of Physics, Beijing Institute of Technology, Beijing, 100081, China}

\author{Yugui Yao}
\email{ygyao@bit.edu.cn}
\affiliation{Centre for Quantum Physics, Key Laboratory of Advanced Optoelectronic Quantum Architecture and Measurement(MOE),
School of Physics, Beijing Institute of Technology, Beijing, 100081, China}
\affiliation{Beijing Key Lab of Nanophotonics $\&$ Ultrafine Optoelectronic Systems,
School of Physics, Beijing Institute of Technology, Beijing, 100081, China}

\begin{abstract}
Topological insulators(TIs) hold the promise as a platform for spintronics applications due to the fascinating spin-momentum locking (SML) of the surface states. One particular interest lies on using TIs as spin polarized sources for spintronics structures. Here, we propose the band tilt-induced nonlinear Nernst effect (NLNE) in TIs as a clean and efficient route to generate high-performance spin polarization (SP). We show in the presence of SML and Hexagonal warping effect, a finite band tilt can induce an imbalance of two spin carriers and effectively give rise to spin-polarized NLNE current in TIs. In our scheme, both the spin current and charge current regime can be achieved under the thermal drive. The obtainable SP can be efficiently tuned either in a smooth or rapid way, exhibiting highly flexible tunability. In addition, near-unity SP can be generated within a wide range of tunable parameters, which is also found to be robust against temperature. Therefore, our work provides a mechanism to realize controllable room-temperature high-degree SP based on TIs, being of essential importance for future spintronics applications.

\end{abstract} 
%\pacs{}

\maketitle %\section{Introduction}\paragraph{\textcolor{blue}{Introduction.\textemdash{}}}

\textcolor{blue}{\it Introduction.\textemdash{}}Topological spintronics which revolves around the topological states and carriers' spin, in addition to carriers' charge, has attracted growing attention over the past decade owing to its promising applications in spin-based future technologies~\cite{spintronics_2004_PRM,semicond_2007np_awschalom,afm_spin_2016_Nat.,afm_spin_2018_PRM, afm_fm_sTorque_2019_Jario, topo_spintronics_2019,review_2020jmmm_hirohata,vandw_2021_nn_sierra,ferrimagnetic_2022nm_kim,TIs_spintronics_2022NM_Wang}. A primary requirement for the practical applications of spintronics is to generate spin polarized carriers utilizing proper materials. Though much progress has been made~\cite{nonMag_sp_2004_prl,nonMag_spinValley_2012_prl,hiddenspin_2014_np, noncollinear_yan_2017,TI_spincurrent_2017,ferro_nanowire_2019,vdw_2020_nc,afm_2020_nc,noncollinear_2020_prb, fm_2020_plA,afm_sSplitter_2021_Jario,optospin_2021, spinPVE_2021_nc,2022_nlshe_Hayami}, the realization of high-performance spin-polarized currents, i.e., with high degree of SP, long lifetime and robustness against temperature, remains to date a critical challenge in spintronics.

3D TIs are inherent SP generators~\cite{TIs_spintronics_Wang2016,TIs_techreview_2021}, which allows for generating SP without the need of magnetism such as ferromagnetic layers. Several experiments have demonstrated the observation of the (electrical) current-induced SP in TIs~\cite{sp_detection_2014_Nnanotech,sp_detect_2014_nanolett,sp_transport_2014_ssc,sp_detection_2015_sr,sp_current_2015_nanolett,sp_tunneling_2015_prb,sp_mapping_2015_prb, TI_spincurrent_2017, sp_comparison_2016_nc}, which are believed to be resulted from the intrinsic SML of the topological surface states (TSSs). Besides SML, however, it is reported that bulk spin Hall effect as well as the trivial 2D electron gas with strong spin-orbit coupling near the surfaces in TIs are also possible contributors~\cite{sp_comparison_2016_nc,nonMag_sp_2016_nc,2dEG_2011prl_king,2dEG_2012nc_Bahramy,origin_ciSP_2015prb_Banerjee,othermechanism_2021prb_Tian,ciSP_2021prb_xueqikun}. This makes the origin of the experimentally-measured SP in TIs unclear. While the search for high-quality TIs that can reach bulk-insulating states will exclude most of the disruptive SP contributors, it can not guarantee the SP generated in TIs currently are applicable for spintronics. As has been pointed out recently~\cite{SPinterpreting_2016prb_Li}, the net SP generated in these experiments are mostly contributions from the in-plane spin projection of the intrinsic states, rendering the nonequilibrium ensemble SP obatinable in TIs being relatively minuscule. Consequently, to utilize TIs as spin polarized sources, alternative mechanisms for the SP generation are still highly demanded. 

\begin{figure}[t!]
    \begin{centering}
       \vspace{-4mm}
        \includegraphics[width=0.48\textwidth]{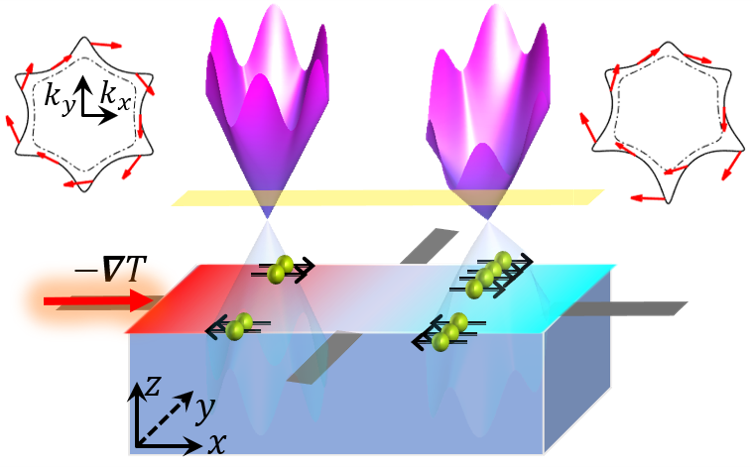}
        \par\end{centering}
        \vspace{-4mm}
        \centering{}\caption{Schematic experimental setup for measuring the spin-polarized NLNE in 3D TIs. A transverse charge current can be probed from the two leads (in grey) along $y$-direction as a second-order response to the longitudinal heat current flowing on the top surface. The left Dirac cone is non-tilted while the right one is slightly tilted along $k_y$-direction. The warped contour plots of a given Fermi surface (yellow plane) are also presented.%, where the red arrows and dashed lines represent for the spin textures and the Fermi surfaces with particle-hole asymmetry, respectively. 
        } \label{FIG.1} 
        \vspace{-4mm}
\end{figure}

Despite the intense interest and early success in the filed of nonlinear responses~\cite{nlPerspective_2021_Xie}, transport effects in the respect of SP (or spin-polarized transports), have not yet been discussed in the nonlinear regime so far. Here, we propose the band-tilt induced NLNE in 3D TIs as a new mechanism to generate highly tunable SP. Contrast to the NLNEs reported previously~\cite{nlNernst_2019_zeng,nlNernst_2019_xqy,BCD_Weyl_2021_Zeng,nlNernst_magnus_2020,nl_spinNernst_2021akagi,chiral_weyl_2020_zeng,Yu_nlpNernst_2021}, 
%the one investigated here neither involves the Berry curvature (dipole)~\cite{nlNernst_2019_zeng,nlNernst_2019_xqy,BCD_Weyl_2021_Zeng,nlNernst_magnus_2020,nl_spinNernst_2021akagi} or chiral anomaly effect~\cite{chiral_weyl_2020_zeng}, nor requires in-plane magnetic field to interconvert the nonlinear spin and charge current~\cite{Yu_nlpNernst_2021,NLPHE_2019_prl}. Instead, it 
the one investigated here has a quantum origin from the finite band tilt effect in TIs, which effectively generates the transverse nonlinear spin and charge currents coplanar with the thermal gradient. Interestingly, accompanied with the SML and hexagonal warping effect in TIs, the band tilt-induced distortion of the Fermi surface results in an imbalance of spin-up and spin-down carriers, as schematically shown in Fig.~\ref{FIG.1}. As an immediate result, nonzero net SP dependent on multi parameters can be generated under a thermal drive. Being a cooperative effect of SML, hexagonal warping, and band tilt, the thermally induced SP carried by NLNE current will not be concealed by any other possible states, in striking contrast to the previous SP generations in TIs. %that is merely ascribed to SML in TIs.
Moreover, since the generated SP is associated with the NLNE, which is a purely nonequilibrium Fermi surface quantity~\cite{nlPerspective_2021_Xie,Yu_nlpNernst_2021,NLPHE_2019_prl}, it represents an ensemble SP density exactly caused by momentum asymmetry. Thus the obtainable SP is a pure nonequilibrium ensemble quantity and expected to be fully applicable for spintronics. We further show the proposed NLNE in TIs, remarkably, provides near-unity SP with high tunability as well as robustness against temperature.  
Therefore, our scheme to generate SP will pave the way for TI-based spintronics applications. In what follows, we present the semiclassical formalism for the spin-polarized NLNE in 3D TIs and demonstrate the generation of high-performance SP in detail.

%%%%%%%%%%%%%%%%%%%%%%%%%%%%%%%%%%%%%%%%%%%%%%%%%%%%%%%%%%%
%%%%%%%%%%%%%%%%%%%%%%%%%%%%%%%%%%%%%%%%%%%%%%%%%%%%%%%

\textcolor{blue}{\it Surface states in 3D TIs.\textemdash{}}The standard Hamiltonian for the surface state of TI is given by
\begin{equation}
    H_0 = \Big[v_{F}\hbar \bm{k \times \hat{z}} + \lambda \bm{k\times \hat{y}}(k^2_x-3 k^2_y)\Big]\bm{\sigma}  ,\label{Eq:Hamiltonian_0}
\end{equation} 
where $v_f$ is the Dirac velocity and $\lambda$ denotes the hexagonal warping strength promised by the $C_{3v}$ rotational invariance the rhombohedral crystal structure (e.g. $\mathrm{Bi_2 Te_3}$)~\cite{Liang_2009_warping}. It has been shown by \textit{Zhang et al.}~\cite{tilt_TSS_2012_zhang,breakingTRS_Zhang_2013PRL}, that the TSSs can be efficiently manipulated through adjusting the surface potential, such as introducing the in-plane (surface) magnetization via doping the TIs or the proximity effect of ferromagnetic insulators. In such cases, both $C_{3v}$ and time-reversal symmetry (TRS) of the system would be broken, and the surface state Hamiltonian will be correspondingly modified. To the leading term, it can be effectively described by the band tilt $H_t = v_t k_t \sigma_0$ ($t=x, y$). Therefore, an effective TSS Hamiltonian can be written as
\begin{equation}
    H =H_0 + H_t.\label{Eq:Hamiltonian}
\end{equation}
In effect, besides the linear band tilt term, the symmetry breaking can also lead to additional terms (higher-order in $\bm{k}$) in the surface Hamiltonian, which might also be important for the physical properties of TSSs. As a conceptual and theoretical work, here we mainly focus on the tilt-induced transport effects for the TSSs.
%\textcolor{red}{Note that, the band tilt included above naturally exists when both time-reversal-symmetry (TRS) and $C_{3v}$ are broken for the surface states of TIs.} %Such a tilting effect, as we will show follows, is crucial to generate the SP and NLNE. 

The surface band dispersion based on Eq.~(\ref{Eq:Hamiltonian}) is 
$E_{\pm, \bm{k}} = v_t k_t \pm \sqrt{(v_f \hbar)^2k^2+\lambda ^2 k^6 \cos^2{3\phi}}$, where $\phi$ is the azimuth angle of momentum $\bm{k}$ with respect to $k_x$-axis, ``$\pm$" denotes the conduction (upper) and valence (lower) band, respectively, and the corresponding eigenstates are defined as $|\pm, \bm{k}\>$. Note that, the warping term coupled to $\sigma_z$ in Eq.~(\ref{Eq:Hamiltonian}) explicitly introduces an out-of-plane SP along $s_z$ direction~\cite{bimer_2018_np}. However, here we are more interested in the in-plane SP $\<\sigma_{||}\>$, obtained as $\<\pm, \bm{k}|\sigma_{||}|\pm, \bm{k}\> =(\<\sigma_x\>, \<\sigma_y\>) =\pm \frac{v_f \hbar}{E_0} (k_y, -k_x)$ with $E_0 =\big|E_{\pm, \bm{k}}\big|_{D=0, v_t=0}$. This explicitly reveals the SP is perpendicularly locked to the momentum, independent of the warping and/or band tilting effect, as shown in Fig.~\ref{FIG.1}.

Before proceeding, we want to mention that a particle-hole asymmetry (PHA) term ($\propto k^2$) can also exist for the TSSs, which has been found important in many intriguing effects in TIs~\cite{NLPHE_2019_prl,Nagaosa_spin_2017,Ani_spin_2019,PHE_tilt_2020,super_transport_2021}. %It has also been considered as an additional contributor to the nonlinear Hall effect recently~\cite{NLPHE_2019_prl}. 
In effect, as shown by the contour plots in black dashed lines in Fig.~\ref{FIG.1}, it suppresses the warping effect, which in turn inevitably weakens the warping-effect-dependent nonlinear responses. Fortunately, it can be found that in $\mathrm{Bi_2Te_3}$ the warping and tilting effect is just slightly suppressed by PHA, especially at relatively lower Fermi energies, %than in $\mathrm{Bi_2Se_3}$, 
which thus will not affect the main results of this work~\cite{SM}. 
%the nonlinear effects remain experimentally measurable~\cite{SM}. 
To better elaborate the generation mechanism of the spin polarized NLNE, we here in the main text consider the particle-hole symmetric TSSs for the following discussions.

\textcolor{blue}{\it Semiclassical Boltzmann transport formalization.\textemdash{}}The nonequilibrium distribution of carriers in response to external fields can be phenomenally depicted by the Boltzmann transport theory. Within the relaxation time approximation, the Boltzmann transport equation in steady state is 
\begin{equation}
\tau (\bm{\dot{r}} \partial_r f +\bm{\dot{k}} \partial_k f) = -(f-f_0), \label{Eq:BTE}
\end{equation}
where $f_0$ and $f$ is the equilibrium and  perturbed Fermi-Dirac distribution function, respectively. For simplicity, the relaxation time $\tau$ will be treated as a constant. 
%Here, $\tau$ denotes the relaxation time dependent on the detailed scattering processes. For simplicity, we will treat it as a constant in the following discussions. 

Besides electrical generation, applying thermal gradient is another feasible approach to generate SP~\cite{thermal_2010sscommu_WANG,fm_spinInjection_2010_Nat.p_junction,fm_spinC_2011_Nat_junction,thermal_2013prb_dyrdal,thermal_2016frontiers_zhongshui,thermal_2018prb_dyrdal}. Specifically, the linear thermoelectric effect has been recently reported to generate and control the spin-polarized current in topological materials~\cite{spc_2021_prb}. Here let's consider a configuration in which only the thermal gradient is present, i.e., $\bm{E}, \bm{B}=0, \bm{\nabla }T \neq 0$. Thus, Eq.~(\ref{Eq:BTE}) can be reduced to $\tau \bm{v} \cdot \partial_{\bm{r}} f  = \sum_{n} g^{(n)}_{\bm{k}}$, where $g^{(n)}_{\bm{k}}$ implies the $n_{th}$-order ($n \geq 1$) correction in $\bm{\nabla} T$ to $f_0$, i.e., $f =f_0 + \sum_{n} g^{(n)}_{\bm{k}}$. The corrections $g^{(n)}_{\bm{k}}$ are generally obtained via the iterative substitutions in Boltzmann equation~\cite{iteration_BTE_2001_prb,SM}, through which we have
$g^{(1)}_{\bm{k}}=\tau \frac{(\e_{\bm{k}}-\mu) }{T\hbar} \frac{\partial f_0}{\partial k_x}  \nabla_x T$ and $g^{(2)}_{\bm{k}} = \tau^2 [ 2 v_x \frac{\partial f_0}{\partial k_x}  + (\e_{\bm{k}} -\mu) \frac{\partial f_0^2}{\partial k^2_x}] \frac{(\e_{\bm{k}}-\mu)}{\hbar^2 T^2}(\nabla_x T)^2$. Without loss of generality, a thermal gradient applied along $x$-direction is considered above. Note that, here the first-order correction $g^{(1)}_{\bm{k}}$ generally leads us to the linear responses, while the correction $g^{(2)}_{\bm{k}}$ is the one plays the key role in the nonlinear transport discussed in the following. 
Analogous to the charge and energy current \cite{Beffect_2010_niu,heatcurrent_2011_niu}, the spin current operator can be defined as $\hat{\bm{j}}^{\bm{s}_{\nu {'}}}_{\nu} =\frac{\hbar}{4} \big\{ \hat{\bm{v}}_{\nu}, \bm{\sigma}_{\nu'}\big\}$, 
%~\cite{scurrent_2004_prl,scurrent_2004_sci}, 
where $\hat{\bm{v}}_{\nu}=\partial H_{\bm{k}}/\hbar \partial \bm{k}_{\nu}$, and $\bm{s}_{\nu'}$ indicates the $\nu'$-spin component with Pauli matrix $\bm{\sigma}_{\nu'}$. In the absence of electric and magnetic field, the semiclassical equations of motion of Bloch electrons are simplified as $\bm{\dot{k}}=0,~\bm{\dot{r}}=\<\hat{\bm{v}}\>_{n}= \frac{\partial \e_{n, \bm{k}}}{\hbar \partial \bm{k}}$ ~\cite{Beffect_2010_niu}. Therefore, the charge current and the spin current are respectively given as, 
\begin{equation}\begin{split}
\bm{j}_{\nu} &= \sum_n \int [d\bm{k}]f_{n, \bm{k}} \<-e\hat{\bm{v}}_{\nu}\>_n , \\
    \bm{j}^{\bm{s}_{\nu'}}_{\nu} &= \sum_n \int [d\bm{k}] f_{n,\bm{k}} \<\hat{\bm{j}}^{\bm{s}_{\nu'}}_{\nu}\>_n,    
\end{split} \label{Eq:3}
\end{equation}
where $f_{n, \bm{k}}$ and $\< \dots \>_n$ represents the distribution function and the average with regard to $n_{th}$ eigenstate $\e_{n, \bm{k}}$, respectively. Note that, $ \bm{j}^{\bm{s}_{\nu'}}_{\nu}$ implies the spin current along $\nu$-direction with spin pointing in direction $\nu'$. In this work, we focus on the nonlinear responses to temperature gradient stemming from the second-order correction $g^{(2)}_{\bm{k}}$. It is important to note that, %for each transport electron, the time-reversal-symmetry (TRS) partner carries an opposite spin as well as opposite velocity, while the inversion-symmetry (IS) partner carries an opposite velocity but the same spin. Therefore, 
the second-order nonlinear spin current only requires broken inversion-symmetry (IS), 
%as a second-order response to temperature gradient can occur in IS broken but time-reversal-invariant materials, 
whereas both the breaking of IS and TRS are required to guarantee a finite second-order nonlinear charge current.

%%%%%%%%%%%%%%%%%%%%%%%%%%%%%%%%%%%%%%%%%%%%%%%%%%%%%%%%%%%%%
\begin{figure}[t!]
    \begin{centering}
        \includegraphics[width=0.46\textwidth]{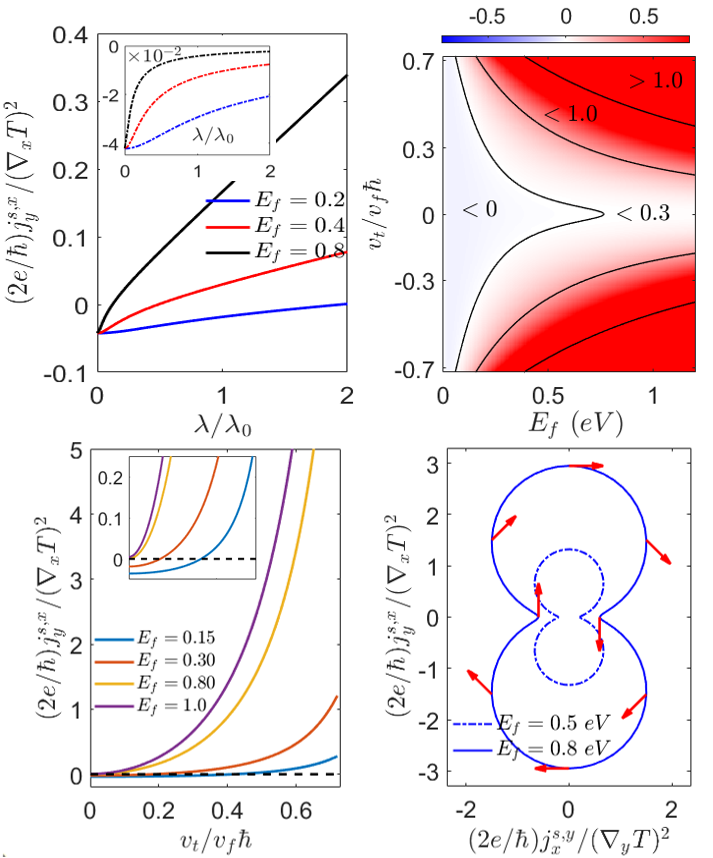}
            \llap{\parbox[b]{163mm}{\large\text{(a)}\\\rule{0ex}{92mm}}}
	\llap{\parbox[b]{82mm}{\large\text{(b)}\\\rule{0ex}{91mm}}}
	\llap{\parbox[b]{164mm}{\large\text{(c)}\\\rule{0ex}{45mm}}}
	\llap{\parbox[b]{84mm}{\large\text{(d)}\\\rule{0ex}{45mm}}}
        \par\end{centering}
        \centering{}
        \vspace{-2mm}
        \caption{%Band tilt enhanced nonlinear spin Nernst effect (NLSNE). 
        (a) NLSNE plotted as a function of the warping strength $\lambda$ at different chemical potentials. The band tilt parameter is $ v_t =0.2 v_{f}\hbar$ while $0$ for the inset. (b) NLSNE as function of Fermi energy $E_{f}$ and band tilting strength $v_{t}$. Here $\lambda_0=250~eV \angstrom^3$ is fixed. (c) NLSNE versus $v_{t}/v_{f}\hbar$ for different Fermi energies. The inset presents the zoomed view around the black dashed zero-line. (d) Texture of the spin current $|j^s_{\theta}|$ in the polar coordinate plane. 
        %The solid and dashed lines represent $|j^s_{\theta}|$ at $E_f=0.8$, $0.5~eV$, respectively, and the red arrows indicate the spin direction of the NLSNE. 
        The other parameters used here are $v_f\hbar =2.25~eV \angstrom, T=50K$.  
        } \label{FIG.2} 
        \vspace{-2mm}
\end{figure}
%%%%%%%%%%%%%%%%%%%%%%%%%%%%%%%%%%%%%%%%%%%%%%%%%%%%%%%%%%%%%%%

\textcolor{blue}{\it Band tilt-enhanced spin current.\textemdash{}} Using the SP obtained for Eq.~(\ref{Eq:Hamiltonian}), the second-order nonlinear spin Nernst current (NLSNE) is found to be
\begin{equation}\begin{split}
    j^{s_x}_{y} & = \tau^2\frac{\hbar}{2} \int [d\bm{k}] \frac{v_f \hbar k_y}{E_0} v_y \bigg[ 2 \hbar v_x \frac{\partial f_0}{\partial k_x}  \\
     &+ (E_{\bm{k}} -\mu) \frac{\partial f_0^2}{\partial k^2_x}\bigg] \frac{(E_{\bm{k}}-\mu)}{\hbar^2 T^2}(\nabla_x T)^2,  \label{Eq:7}   
\end{split}\end{equation}
where the upper branch of the surface state (denoted as $E_{\bm{k}}$) is considered. Note that, $j^{s_x}_{y}$ is finite even without the breaking of additional symmetry, or considering the band warping/band tilting effect. Such a feature is implied by the non-zero intersections at $\lambda =0$ in Fig.~\ref{FIG.2}(a). In addition to the hexagonal warping-induced enhancement~\cite{Yu_nlpNernst_2021}, it shows the NLSNE can be further increased by a finite band tilt [Fig.~\ref{FIG.2}(a)], in contrast to the previously reported nonlinear planar effects that remains basically unchanged even with varying magnetic fields~\cite{NLPHE_2019_prl, Yu_nlpNernst_2021}. Such evident enhancement along with the $v_t$ and $E_f$-dependency are presented in Fig.~\ref{FIG.2}(b), where the colors indicates the magnitudes and the black lines are shown for isovalues at $0, 0.3, 1.0$, respectively. Line cuts with different chemical potentials $E_f$ is also given in Fig.~\ref{FIG.2}(c). Obviously, even a moderate tilting strength ($v_t/v_f\hbar \sim 0.4$) can easily increase the NLSNE by hundreds of times on its order of magnitude compared to the non-tilted case at $v_t=0$, especially at higher chemical potentials.  

One finds that, under the drive of thermal gradient $\nabla_{c}T$, only the NLSNE current with transverse SP ($j^{s_{c {\perp}}}_{c}$, $j_{c_{\perp}}^{s_{c}}$) can exist~\cite{SM}. Here we focus on $j^{s_{c {\perp}}}_{c}$, the component transverse to $\nabla_{c}T$. Particularly, both $j^{s_x}_{y} \propto (\nabla_x T)^2$ and $j^{s_y}_{x} \propto (\nabla_y T)^2$ will be simultaneously generated, given a misalignment between the externally applied thermal gradient and the principal axes. In such case, a real-space spin texture of the generated NLSNE can be determined as  $|j^{s}_{\theta}|=\sqrt{(j^{s_x}_y \sin{\theta})^2 +(j^{s_y}_{x} \cos{\theta})^2}$, where $\theta$ stands for the polar angle in the rotational coordinate system that is commonly exploited in the conventional transport measurement~\cite{nlPerspective_2021_Xie}. The numerically calculated spin current textures $|j^s_{\theta}|$ with $E_f=0.8~eV$ (solid) and $E_f=0.5~eV$ (dashed) are plotted in Fig.~\ref{FIG.2}(d). 
%where the red arrows representing the spin direction. 
Interestingly, the SML texture is still preserved for the NLSNE, irrespective of the anisotropy in magnitude. Another interesting angular dependency may come from the crossed thermal gradient and band tilt~\cite{SM}.
%More discussions about the angular dependency, e.g. caused by misalignment between the thermal gradient and band tilt, of different components of the NLSNE (NLNE) can be similarly obtained~\cite{SM}. 

\textcolor{blue}{\it Nernst current and the spin polarization.\textemdash{}}Based on Eq.~(\ref{Eq:3}), the second-order NLNE is obtained as
\begin{equation}\begin{split}
    j_{y} &= -e \tau^2 \int [d\bm{k}] v_y \bigg[ 2 \hbar v_x \frac{\partial f_0}{\partial k_x}  \\
     &+ (E_{\bm{k}} -\mu) \frac{\partial f_0^2}{\partial k^2_x}\bigg] \frac{(E_{\bm{k}}-\mu)}{\hbar^2 T^2}(\nabla_x T)^2 , \label{Eq:8}
\end{split}\end{equation}
which differs from the NLSNE in Eq.~(\ref{Eq:7}) merely by factor $-\frac{\hbar}{2e}\<\sigma_x\>$ inside the integral. It is worthy noting that $j_y$ vanishes unless $v_y$ contains an extra momentum-independent contribution from the finite band tilt, i.e., $v_y  \rightarrow v_t/\hbar \pm v_y$. As such, we find the generated NLNE current only flows along the direction of the band tilt, e.g. $y$-direction ($\perp \nabla_x T$ ) in the current configuration. This is different from NLSNE which are generated in both transverse and longitude directions. In effect, a longitudinal or transverse nonlinear charge current can be generated if the Dirac cone state is perfectly tilted along $x$- or $y$-direction, respectively~\cite{SM}. 
%To recap, there is no NLNE current generated (namely all the nonlinear currents will flow longitudinally) if the thermal gradient is colinear with the band tilt; while the NLNE signal reaches its maximum when the applied thermal gradient is perfectly perpendicular to the band tilt~\cite{SM}. 

\begin{figure}[t!]
    \begin{centering}
        \includegraphics[width=0.46\textwidth]{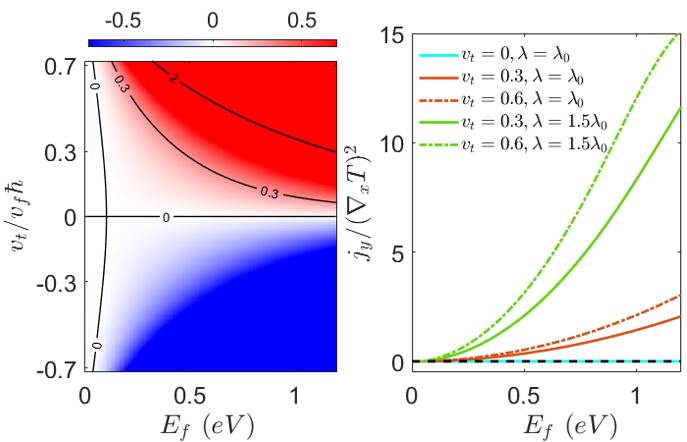}
        \llap{\parbox[b]{158mm}{\large\text{(a)}\\\rule{0ex}{48mm}}}
	\llap{\parbox[b]{82mm}{\large\text{(b)}\\\rule{0ex}{48mm}}}
        \par\end{centering}
        \vspace{-2mm}
        \centering{}\caption{%Nonlinear Nernst effect (NLNE) plotted as function of different parameters.
        The NLNE current $[j_y/(\nabla_x T)^2]$ (a) as function of Fermi energy $E_{f}$ and band tilting strength $v_{t}$ and (b) versus Fermi energy $E_{f}$ at different tilting strength $v_{t}$. $\lambda=\lambda_{0}$ is fixed in (a) and the NLNE current is in units of $nA \cdot\mu m/K^{2}$. The other parameters used here are same to Fig.~\ref{FIG.2}.} \label{FIG.3} 
\vspace{-2mm}
\end{figure}

Unlike the NLSNE, the NLNE $j_{y}$ is asymmetric with respect to tilting strength $v_{t}$, which undergoes a sign change when modulating $v_{t}$ into $-v_{t}$ [Fig.~\ref{FIG.3}(a)]. As restricted by TRS, NLNE current vanishes [$j_{y}=0$, see Fig.~\ref{FIG.3}] in the absence of band tilt ($v_{t}=0$). Yet, nonzero NLNE current immediately emerges when band tilt is finite. Interestingly, it rapidly increases by enhancing the warping effect and/or band tilting effect, and modulating the Fermi energy can also effectively tune the NLNE. These prominent tunabilities of NLNE are consistent to the NLSNE.
%%%%%%%%%%%%%%%%%%%%%%%%%%%%%%%%%%%%%%%%%%%%%%%%%%%%%%%%%%%%%%%%%%%%%%%%%%%%%%%%%%
%%%%%%%%%%%%%%%%%%%%%%%%%%%%%%%%%%%%%%%%%%%%%

\begin{figure}[t!]
    \begin{centering}
        \includegraphics[width=0.46\textwidth]{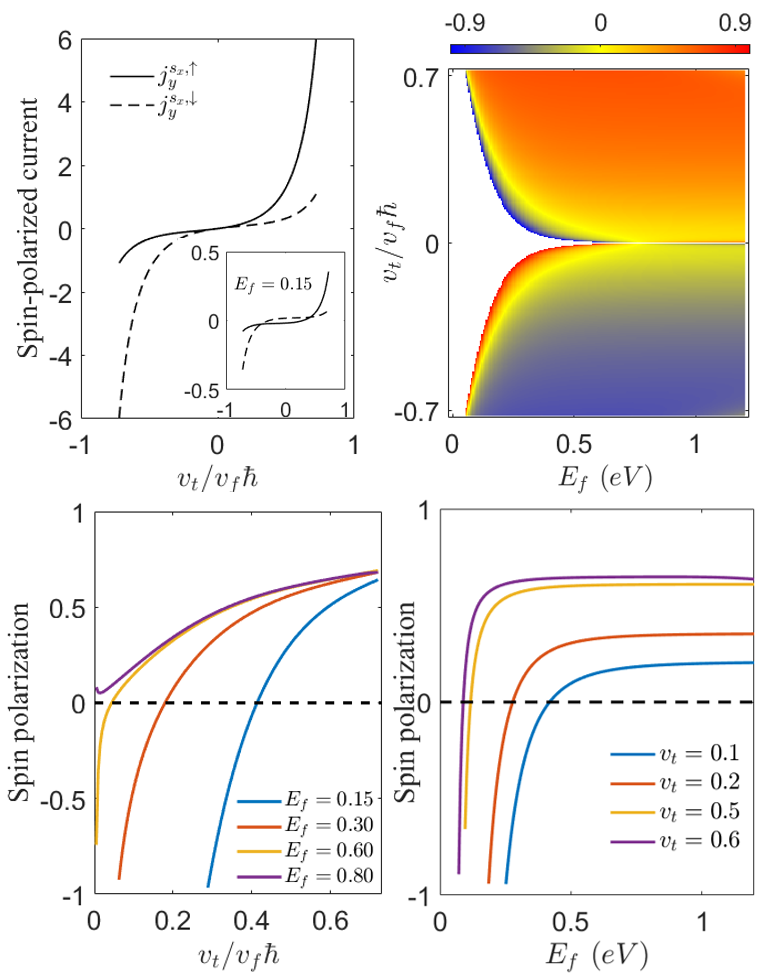}
           \llap{\parbox[b]{163mm}{\large\text{(a)}\\\rule{0ex}{99mm}}}
	\llap{\parbox[b]{84mm}{\large\text{(b)}\\\rule{0ex}{99mm}}}
	\llap{\parbox[b]{162mm}{\large\text{(c)}\\\rule{0ex}{50mm}}}
	\llap{\parbox[b]{85mm}{\large\text{(d)}\\\rule{0ex}{50mm}}}
        \par\end{centering}
        \vspace{-2mm}
        \centering{}\caption{(a) NLNE currents with spin up ($j^{s_x, \uparrow}_{y}$) and spin down ($j^{s_x, \downarrow}_y$)  as a function of band tilting strength $v_{t}$ with the Fermi energy $E_{f}=0.6~eV$, while the inset shows the case of $E_f =0.15~eV$. (b) The SP $\eta= 2ej^{s_{x}}_{y}/\hbar j_{y}$ as a function of Fermi energy $E_{f}$ and band tilting strength $v_{t}$. Panel (c), (d) plot the cut lines taken from panel (b). 
        %with $\eta$ versus $E_{f}$ for different band tilting strength $v_{t}$, while (d) shows $\eta$ versus $v_{t}$ for different Fermi energies.
        } \label{FIG.4} 
        \vspace{-2mm}
\end{figure}

The two opposite spin components can be extracted from the NLNE current, respectively as, $j^{s_x, \uparrow/\downarrow}_y =\frac{1}{2}(j_y \pm \frac{2e}{\hbar}j^{s_x}_y)$, where the coefficient $2e/\hbar$ is incorporated to match the dimension. The above spin-up $ j^{s_x, \uparrow}_y$ and spin-down $ j^{s_x,\downarrow}_y $ component of NLNE current are individually tunable by tilting strength, essentially differing from previous works~\cite{NLPHE_2019_prl, Yu_nlpNernst_2021}. Such tunablity is shown in Fig.~\ref{FIG.4}(a), where $j^{s_x, \uparrow/\downarrow}_y$ at $E_f=0.6~eV$ are plotted as a function of the tilt parameter $v_t$. Interestingly, $ j^{s_x, \uparrow}_y$ and $ j^{s_x,\downarrow}_y $ gets enhanced prominently when the band is tilted along ${y}$-direction ($v_{f}>0$) and negative ${y}$-direction ($v_{f}<0$), respectively. For a given band tilt, an apparent difference in the enhancement on the spin-up and spin-down components is observed. These significant anisotropies in band-tilt dependency (i.e., uneven enhancement on opposite spin) explicitly result in the spin polarized NLNE current (Fig.~\ref{FIG.1}). When fixing $E_f = 0.15~eV$ and varying $v_{t}/v_{f}\hbar$ within $[-0.5, 0.5]$, $j^{s_{x},\downarrow}_{y}$ is observed to be larger than $j^{s_{x},\uparrow}_{y}$ [$j^{s_{x},\downarrow}_{y}> j^{s_{x},\uparrow}_{y}$, inset of  Fig.~\ref{FIG.4}(a)], which leads to a negative NLSNE consistent with Fig.~\ref{FIG.2}(c). Specifically, in the case of $j^{s_{x},\downarrow}_{y} = - j^{s_{x},\uparrow}_{y}$, a pure nonlinear spin current is generated. 
%%%%%%%%%%%%%%%%%%%%%%%%%%%%%%%%%%%%%%%%%%%%
%%%%%%%%%%%%%%%%%%%%%%%%%%%%%%%%%%%%%%%%%%%%%%%%%%%%%%%%%%%%%

Generally, the SP $\eta$ ($|\eta|\leq 1$) is defined as the ratio of the subtraction and the addiction of $j^{s_x, \uparrow}_y$ and $j^{s_x, \downarrow}_{y}$, namely $\eta=2e j^{s_x}_y/\hbar j_y$ in our case. As shown in Fig.~\ref{FIG.4}(b), the white region with $|2e j^{s_x}_{y}/\hbar|>|j_y|$, where NLSNE is dominant over NLNE, shall be considered as the spin current regime with an ill-defined SP $\eta$. The previously proposed nonlinear planar effects~\cite{NLPHE_2019_prl, Yu_nlpNernst_2021}, which engender from spin-to-charge conversion under magnetic field with conversion rate less than $100\%$ (i.e., $|j_y| <|2e j^{s_x}_{y}/\hbar|$), exactly belong to such regime. Moreover, the longitudinal nonlinear currents, if generated in our scheme, also belongs to this regime, as the charge current is evidently smaller in magnitude than the spin current~\cite{SM}. 

For the sake of practical applications, the charge current regime with $|j_y| > |2e j^{s_x}_y/\hbar|$ that corresponds to the colorful regions in Fig.~\ref{FIG.4}(b), is more interesting to us. Note that, the condition for this regime also implies $j^{s_x, \uparrow}_y j^{s_x, \downarrow}_y >0$, namely the spin-up and spin-down components of NLNE are expected to flow in the same direction. This also determines a series of threshold strength of the band tilt~\cite{SM} for generating SP. Remarkably, the SP exhibits considerably high tunabilities in terms of the band tilt and Fermi energy. It undergoes a sign change near the spin balance region that satisfies $j^{s_{x},\uparrow}_{y}=j^{s_{x},\downarrow}_{y}\neq0$ or $\eta=0$, when varying band tilting strength $v_{t}$ with fixed Fermi energy or modulating Fermi energy with a selected $v_{t}$ [ Fig.~\ref{FIG.4}(b)]. This interesting feature may promise the realization of sign-switchable SP, which is also significant for spintronics applications. Moreover, SP varies rapidly before the spin balance while gradually tends to be a constant after going through it [Fig.~\ref{FIG.4}(c), (d)], offering a rapid and smooth tuning regime respectively. Such tunability (sensitiveness) difference is especially evident for the case of a fixed tilt but varying chemical potential [Fig.~\ref{FIG.4}(d)]. In addition, it is also shown that NLNE current with near unity ($\sim100\%$) SP can be generated in a wide range of Fermi energy and band tilting strength [see the dark blue and dark red colors in Fig.~\ref{FIG.4}(b)]. 

The spin components involved in the nonzero spin-polarized NLNE always remain orthogonal to the charge and spin flows [see Eqs.~(\ref{Eq:7}, \ref{Eq:8})], which enables the generation of perfect transverse SP with high performance. Since both the NLSNE and NLNE are nearly unchanged in the presence of a finite magnetic field~\cite{SM}, the SP generation in our scheme cannot be affected easily by the external magnetic field. Additionally, the impact of varying temperature on our proposed spin-polarized NLNE is found to be negligible (see SM~\cite{SM}), consistent with the nonlinear planar Nernst effect studied in Ref.~\cite{Yu_nlpNernst_2021}. Thus our scheme can also achieve robust room-temperature SP, 
%As noted earlier, the obtained SP in our scheme also exhibit highly flexible tunability over multi controllable parameters, through which, near-unity degree of polarization can be easily reached, all 
being of essential importance in the spintronics applications. %We also find that, the proposed features of the spin-polarized NLNE remain invariant even in the presence of magnetic field. 

\textcolor{blue}{\it Discussion and conclusion\textemdash{}}
The proposed spin-polarized NLNE in this work can be ascribed to the joint effect of SML, hexagonal warping and band tilting effect, where the former two ingredients are quite generic in realistic materials. The tilt of Dirac cone can be also realized in various materials and manifests interesting effects on the low-energy behaviors of Dirac fermions~\cite{interplay_2018,interplay_2019,PHE_tilt_2020, super_transport_2021,tilted_graphene_2010,tilted_DSM_graphene_2008, TCI_2012,TCI_Inti_2015,TCI_2017}. For TSSs in TIs, the finite band tilt is symmetry-allowed (or naturally exist) when TRS and $C_{3v}$ are both broken, as pointed out earlier in this work. %e.g., through in-plane magnetization doping or proximity effect of ferromagnetic insulators~\cite{tilt_TSS_2012_zhang,breakingTRS_Zhang_2013PRL}. 
Additionally, the magnetic topological semimetals that hold tilted Dirac cones, provided IS is further broken, can also support nonzero spin-polarized NLNE in principle. 

The spin-polarized NLNE can be quantitatively estimated via the charge current $I = j_y \times l$ with $l$ the length of the material. Using the model parameters fitted for the TSSs of realistic material $\mathrm{Bi_2Te_3}$~\cite{Liang_2009_warping,2010prb_fpstudies_cxLiu,2011prb_nonidealcone_YOichi, 2009Sci_shen,2009Np_exp,tss_2010prl,Ando_2014prb,kimura_2018prb}, estimations of $I\sim 0.03~\mu A, \eta \sim 0.30$ can be obtained for a film with $ l\sim100~\mu m$~\cite{NLPHE_2019_prl}, when Fermi energy $E_f =0.49~eV$, in-plane thermal gradient $\nabla_x T \sim-1.3~K\mu m^{-1}$~\cite{2019NanoLett_DeltaT} and a moderate band tilt $v_t/v_f\hbar = 0.2$ are considered. Remarkably, the higher SP and/or charge current magnitude can be realized by tuning different parameters. For instance, a much higher SP $\eta \sim0.965$ carried by the NLNE can be achieved at relatively lower Fermi energy $E_f =0.21 eV$, and when the strength of the band tilt is further increased e.g., $v_t/v_f\hbar =0.45$, a spin-polarized current $I$ with magnitude around $0.12~\mu A$ can be realized. The above estimations are of the similar order of the magnitude as that made in the recent work~\cite{Yu_nlpNernst_2021}, which are accessible in relevant experiment.

Though the top and bottom surfaces contribute oppositely to the surface transport effects, it is general that one of the surfaces is dominant as demonstrated in previous studies~\cite{2010Nanolett_Pablo,2012NC_cui,2015NC_Tokura,prl2016_Tokura,bimer_2018_np}, rendering the total responses to be nonzero. Therefore, our model analysis involving only the top surface state is sufficient to capture the essential feature of the proposed spin-polarized nonlinear effect. 
%\textcolor{red}{Using model parameters of the realistic material $\mathrm{Bi_2Te_3}$, numerical estimations of the spin-polarized NLNE are also made for experimental measurement~\cite{SM}}. 
It should be mentioned that the introduced mechanism to generate SP by our work can be generalized to other transport effects (e.g., the nonlinear Hall effect and thermal Hall effect) in other potential materials e.g., the trigonally warped graphene and bilayer graphene~\cite{nl_opto_granphene_2014,Dashui_graphene_warping_2018, BCD_graphene}, etc. Besides, to better reveal the tilt-induced spin-polarized NLNE, a more detailed analysis of the scattering processes that especially incorporate magnetic disorders will be necessarily important. It is also possible in principle to realize the alternating (AC) or second-harmonic-type SP through our mechanism, provided the external driving force (thermal gradient or electric field) is time (frequency)-dependent. These interesting topics beyond the scope of the current work are all left for future studies.  

In conclusion, we successfully demonstrate a nonlinear transport effect
%the band tilt-induced spin-polarized NLNE in the complete absence of the magnetic field and Berry curvature-involved effects. Such a novel transport effect 
that can remarkably realize the high-degree (near-unity) SP with flexible parameter tunabilities as well as robustness against temperature, which thereby hints significant advance and possible applications in spintronics.
%TI-based spintronics.
% With these advantageous properties mentioned above, the spin-polarized NLNE proposed here is thus definitely of significiant importance for the generation and manipulation of the SP for the future applications of spintronics.

% feasible alternative 
\textcolor{blue}{\it Acknowledgments.\textemdash{}}The work is supported by the National Key R\&D Program of China (Grant No. 2020YFA0308800), the NSF of China (Grants Nos. 11734003, 12061131002, 12104043)), the Strategic Priority Research Program of Chinese Academy of Sciences (Grant No. XDB30000000), and the fellowship of the China Postdoctoral Science Foundation (Grant No. 2021M690409). X. Q. Y is supported by the Fundamental Research Funds for the Central Universities and the NSFC (Grant No. 12004107).

\bibliography{my}

\begin{thebibliography}{94}%
\makeatletter
\providecommand \@ifxundefined [1]{%
 \@ifx{#1\undefined}
}%
\providecommand \@ifnum [1]{%
 \ifnum #1\expandafter \@firstoftwo
 \else \expandafter \@secondoftwo
 \fi
}%
\providecommand \@ifx [1]{%
 \ifx #1\expandafter \@firstoftwo
 \else \expandafter \@secondoftwo
 \fi
}%
\providecommand \natexlab [1]{#1}%
\providecommand \enquote  [1]{``#1''}%
\providecommand \bibnamefont  [1]{#1}%
\providecommand \bibfnamefont [1]{#1}%
\providecommand \citenamefont [1]{#1}%
\providecommand \href@noop [0]{\@secondoftwo}%
\providecommand \href [0]{\begingroup \@sanitize@url \@href}%
\providecommand \@href[1]{\@@startlink{#1}\@@href}%
\providecommand \@@href[1]{\endgroup#1\@@endlink}%
\providecommand \@sanitize@url [0]{\catcode `\\12\catcode `\$12\catcode
  `\&12\catcode `\#12\catcode `\^12\catcode `\_12\catcode `\%12\relax}%
\providecommand \@@startlink[1]{}%
\providecommand \@@endlink[0]{}%
\providecommand \url  [0]{\begingroup\@sanitize@url \@url }%
\providecommand \@url [1]{\endgroup\@href {#1}{\urlprefix }}%
\providecommand \urlprefix  [0]{URL }%
\providecommand \Eprint [0]{\href }%
\providecommand \doibase [0]{https://doi.org/}%
\providecommand \selectlanguage [0]{\@gobble}%
\providecommand \bibinfo  [0]{\@secondoftwo}%
\providecommand \bibfield  [0]{\@secondoftwo}%
\providecommand \translation [1]{[#1]}%
\providecommand \BibitemOpen [0]{}%
\providecommand \bibitemStop [0]{}%
\providecommand \bibitemNoStop [0]{.\EOS\space}%
\providecommand \EOS [0]{\spacefactor3000\relax}%
\providecommand \BibitemShut  [1]{\csname bibitem#1\endcsname}%
\let\auto@bib@innerbib\@empty
%</preamble>
\bibitem [{\citenamefont {\ifmmode \check{Z}\else
  \v{Z}\fi{}uti\ifmmode~\acute{c}\else \'{c}\fi{}}\ \emph
  {et~al.}(2004)\citenamefont {\ifmmode \check{Z}\else
  \v{Z}\fi{}uti\ifmmode~\acute{c}\else \'{c}\fi{}}, \citenamefont {Fabian},\
  and\ \citenamefont {Das~Sarma}}]{spintronics_2004_PRM}%
  \BibitemOpen
  \bibfield  {author} {\bibinfo {author} {\bibfnamefont {I.}~\bibnamefont
  {\ifmmode \check{Z}\else \v{Z}\fi{}uti\ifmmode~\acute{c}\else \'{c}\fi{}}},
  \bibinfo {author} {\bibfnamefont {J.}~\bibnamefont {Fabian}},\ and\ \bibinfo
  {author} {\bibfnamefont {S.}~\bibnamefont {Das~Sarma}},\ }\bibfield  {title}
  {\bibinfo {title} {Spintronics: Fundamentals and applications},\ }\href
  {https://doi.org/10.1103/RevModPhys.76.323} {\bibfield  {journal} {\bibinfo
  {journal} {Rev. Mod. Phys.}\ }\textbf {\bibinfo {volume} {76}},\ \bibinfo
  {pages} {323} (\bibinfo {year} {2004})}\BibitemShut {NoStop}%
\bibitem [{\citenamefont {Awschalom}\ and\ \citenamefont
  {Flatt{\'e}}(2007)}]{semicond_2007np_awschalom}%
  \BibitemOpen
  \bibfield  {author} {\bibinfo {author} {\bibfnamefont {D.~D.}\ \bibnamefont
  {Awschalom}}\ and\ \bibinfo {author} {\bibfnamefont {M.~E.}\ \bibnamefont
  {Flatt{\'e}}},\ }\bibfield  {title} {\bibinfo {title} {Challenges for
  semiconductor spintronics},\ }\href@noop {} {\bibfield  {journal} {\bibinfo
  {journal} {Nature physics}\ }\textbf {\bibinfo {volume} {3}},\ \bibinfo
  {pages} {153} (\bibinfo {year} {2007})}\BibitemShut {NoStop}%
\bibitem [{\citenamefont {Jungwirth}\ \emph {et~al.}(2016)\citenamefont
  {Jungwirth}, \citenamefont {Marti}, \citenamefont {Wadley},\ and\
  \citenamefont {Wunderlich}}]{afm_spin_2016_Nat.}%
  \BibitemOpen
  \bibfield  {author} {\bibinfo {author} {\bibfnamefont {T.}~\bibnamefont
  {Jungwirth}}, \bibinfo {author} {\bibfnamefont {X.}~\bibnamefont {Marti}},
  \bibinfo {author} {\bibfnamefont {P.}~\bibnamefont {Wadley}},\ and\ \bibinfo
  {author} {\bibfnamefont {J.}~\bibnamefont {Wunderlich}},\ }\bibfield  {title}
  {\bibinfo {title} {Antiferromagnetic spintronics},\ }\href
  {https://doi.org/10.1038/nnano.2016.18} {\bibfield  {journal} {\bibinfo
  {journal} {Nature Nanotechnology}\ }\textbf {\bibinfo {volume} {11}},\
  \bibinfo {pages} {231–241} (\bibinfo {year} {2016})}\BibitemShut {NoStop}%
\bibitem [{\citenamefont {Baltz}\ \emph {et~al.}(2018)\citenamefont {Baltz},
  \citenamefont {Manchon}, \citenamefont {Tsoi}, \citenamefont {Moriyama},
  \citenamefont {Ono},\ and\ \citenamefont {Tserkovnyak}}]{afm_spin_2018_PRM}%
  \BibitemOpen
  \bibfield  {author} {\bibinfo {author} {\bibfnamefont {V.}~\bibnamefont
  {Baltz}}, \bibinfo {author} {\bibfnamefont {A.}~\bibnamefont {Manchon}},
  \bibinfo {author} {\bibfnamefont {M.}~\bibnamefont {Tsoi}}, \bibinfo {author}
  {\bibfnamefont {T.}~\bibnamefont {Moriyama}}, \bibinfo {author}
  {\bibfnamefont {T.}~\bibnamefont {Ono}},\ and\ \bibinfo {author}
  {\bibfnamefont {Y.}~\bibnamefont {Tserkovnyak}},\ }\bibfield  {title}
  {\bibinfo {title} {Antiferromagnetic spintronics},\ }\href
  {https://doi.org/10.1103/RevModPhys.90.015005} {\bibfield  {journal}
  {\bibinfo  {journal} {Rev. Mod. Phys.}\ }\textbf {\bibinfo {volume} {90}},\
  \bibinfo {pages} {015005} (\bibinfo {year} {2018})}\BibitemShut {NoStop}%
\bibitem [{\citenamefont {Manchon}\ \emph {et~al.}(2019)\citenamefont
  {Manchon}, \citenamefont {\ifmmode~\check{Z}\else \v{Z}\fi{}elezn\'y},
  \citenamefont {Miron}, \citenamefont {Jungwirth}, \citenamefont {Sinova},
  \citenamefont {Thiaville}, \citenamefont {Garello},\ and\ \citenamefont
  {Gambardella}}]{afm_fm_sTorque_2019_Jario}%
  \BibitemOpen
  \bibfield  {author} {\bibinfo {author} {\bibfnamefont {A.}~\bibnamefont
  {Manchon}}, \bibinfo {author} {\bibfnamefont {J.}~\bibnamefont
  {\ifmmode~\check{Z}\else \v{Z}\fi{}elezn\'y}}, \bibinfo {author}
  {\bibfnamefont {I.~M.}\ \bibnamefont {Miron}}, \bibinfo {author}
  {\bibfnamefont {T.}~\bibnamefont {Jungwirth}}, \bibinfo {author}
  {\bibfnamefont {J.}~\bibnamefont {Sinova}}, \bibinfo {author} {\bibfnamefont
  {A.}~\bibnamefont {Thiaville}}, \bibinfo {author} {\bibfnamefont
  {K.}~\bibnamefont {Garello}},\ and\ \bibinfo {author} {\bibfnamefont
  {P.}~\bibnamefont {Gambardella}},\ }\bibfield  {title} {\bibinfo {title}
  {Current-induced spin-orbit torques in ferromagnetic and antiferromagnetic
  systems},\ }\href {https://doi.org/10.1103/RevModPhys.91.035004} {\bibfield
  {journal} {\bibinfo  {journal} {Rev. Mod. Phys.}\ }\textbf {\bibinfo {volume}
  {91}},\ \bibinfo {pages} {035004} (\bibinfo {year} {2019})}\BibitemShut
  {NoStop}%
\bibitem [{\citenamefont {Mengyun~He}(2019)}]{topo_spintronics_2019}%
  \BibitemOpen
  \bibfield  {author} {\bibinfo {author} {\bibfnamefont {Q.~L.~H.}\
  \bibnamefont {Mengyun~He}, \bibfnamefont {Huimin~Sun}},\ }\bibfield  {title}
  {\bibinfo {title} {Topological insulator: Spintronics and quantum
  computations},\ }\href {https://doi.org/10.1007/s11467-019-0893-4} {\bibfield
   {journal} {\bibinfo  {journal} {Frontiers of Physics}\ }\textbf {\bibinfo
  {volume} {14}},\ \bibinfo {eid} {43401} (\bibinfo {year} {2019})}\BibitemShut
  {NoStop}%
\bibitem [{\citenamefont {Hirohata}\ \emph {et~al.}(2020)\citenamefont
  {Hirohata}, \citenamefont {Yamada}, \citenamefont {Nakatani}, \citenamefont
  {Prejbeanu}, \citenamefont {Di{\'e}ny}, \citenamefont {Pirro},\ and\
  \citenamefont {Hillebrands}}]{review_2020jmmm_hirohata}%
  \BibitemOpen
  \bibfield  {author} {\bibinfo {author} {\bibfnamefont {A.}~\bibnamefont
  {Hirohata}}, \bibinfo {author} {\bibfnamefont {K.}~\bibnamefont {Yamada}},
  \bibinfo {author} {\bibfnamefont {Y.}~\bibnamefont {Nakatani}}, \bibinfo
  {author} {\bibfnamefont {I.-L.}\ \bibnamefont {Prejbeanu}}, \bibinfo {author}
  {\bibfnamefont {B.}~\bibnamefont {Di{\'e}ny}}, \bibinfo {author}
  {\bibfnamefont {P.}~\bibnamefont {Pirro}},\ and\ \bibinfo {author}
  {\bibfnamefont {B.}~\bibnamefont {Hillebrands}},\ }\bibfield  {title}
  {\bibinfo {title} {Review on spintronics: Principles and device
  applications},\ }\href@noop {} {\bibfield  {journal} {\bibinfo  {journal}
  {Journal of Magnetism and Magnetic Materials}\ }\textbf {\bibinfo {volume}
  {509}},\ \bibinfo {pages} {166711} (\bibinfo {year} {2020})}\BibitemShut
  {NoStop}%
\bibitem [{\citenamefont {Sierra}\ \emph {et~al.}(2021)\citenamefont {Sierra},
  \citenamefont {Fabian}, \citenamefont {Kawakami}, \citenamefont {Roche},\
  and\ \citenamefont {Valenzuela}}]{vandw_2021_nn_sierra}%
  \BibitemOpen
  \bibfield  {author} {\bibinfo {author} {\bibfnamefont {J.~F.}\ \bibnamefont
  {Sierra}}, \bibinfo {author} {\bibfnamefont {J.}~\bibnamefont {Fabian}},
  \bibinfo {author} {\bibfnamefont {R.~K.}\ \bibnamefont {Kawakami}}, \bibinfo
  {author} {\bibfnamefont {S.}~\bibnamefont {Roche}},\ and\ \bibinfo {author}
  {\bibfnamefont {S.~O.}\ \bibnamefont {Valenzuela}},\ }\bibfield  {title}
  {\bibinfo {title} {Van der waals heterostructures for spintronics and
  opto-spintronics},\ }\href@noop {} {\bibfield  {journal} {\bibinfo  {journal}
  {Nature Nanotechnology}\ }\textbf {\bibinfo {volume} {16}},\ \bibinfo {pages}
  {856} (\bibinfo {year} {2021})}\BibitemShut {NoStop}%
\bibitem [{\citenamefont {Kim}\ \emph {et~al.}(2022)\citenamefont {Kim},
  \citenamefont {Beach}, \citenamefont {Lee}, \citenamefont {Ono},
  \citenamefont {Rasing},\ and\ \citenamefont
  {Yang}}]{ferrimagnetic_2022nm_kim}%
  \BibitemOpen
  \bibfield  {author} {\bibinfo {author} {\bibfnamefont {S.~K.}\ \bibnamefont
  {Kim}}, \bibinfo {author} {\bibfnamefont {G.~S.}\ \bibnamefont {Beach}},
  \bibinfo {author} {\bibfnamefont {K.-J.}\ \bibnamefont {Lee}}, \bibinfo
  {author} {\bibfnamefont {T.}~\bibnamefont {Ono}}, \bibinfo {author}
  {\bibfnamefont {T.}~\bibnamefont {Rasing}},\ and\ \bibinfo {author}
  {\bibfnamefont {H.}~\bibnamefont {Yang}},\ }\bibfield  {title} {\bibinfo
  {title} {Ferrimagnetic spintronics},\ }\href@noop {} {\bibfield  {journal}
  {\bibinfo  {journal} {Nature materials}\ }\textbf {\bibinfo {volume} {21}},\
  \bibinfo {pages} {24} (\bibinfo {year} {2022})}\BibitemShut {NoStop}%
\bibitem [{\citenamefont {{He}}\ \emph {et~al.}(2022)\citenamefont {{He}},
  \citenamefont {{Hughes}}, \citenamefont {{Armitage}}, \citenamefont
  {{Tokura}},\ and\ \citenamefont {{Wang}}}]{TIs_spintronics_2022NM_Wang}%
  \BibitemOpen
  \bibfield  {author} {\bibinfo {author} {\bibfnamefont {Q.~L.}\ \bibnamefont
  {{He}}}, \bibinfo {author} {\bibfnamefont {T.~L.}\ \bibnamefont {{Hughes}}},
  \bibinfo {author} {\bibfnamefont {N.~P.}\ \bibnamefont {{Armitage}}},
  \bibinfo {author} {\bibfnamefont {Y.}~\bibnamefont {{Tokura}}},\ and\
  \bibinfo {author} {\bibfnamefont {K.~L.}\ \bibnamefont {{Wang}}},\ }\bibfield
   {title} {\bibinfo {title} {{Topological spintronics and
  magnetoelectronics}},\ }\href {https://doi.org/10.1038/s41563-021-01138-5}
  {\bibfield  {journal} {\bibinfo  {journal} {Nature Materials}\ }\textbf
  {\bibinfo {volume} {21}},\ \bibinfo {pages} {15} (\bibinfo {year}
  {2022})}\BibitemShut {NoStop}%
\bibitem [{\citenamefont {Khodas}\ \emph {et~al.}(2004)\citenamefont {Khodas},
  \citenamefont {Shekhter},\ and\ \citenamefont
  {Finkel'stein}}]{nonMag_sp_2004_prl}%
  \BibitemOpen
  \bibfield  {author} {\bibinfo {author} {\bibfnamefont {M.}~\bibnamefont
  {Khodas}}, \bibinfo {author} {\bibfnamefont {A.}~\bibnamefont {Shekhter}},\
  and\ \bibinfo {author} {\bibfnamefont {A.~M.}\ \bibnamefont {Finkel'stein}},\
  }\bibfield  {title} {\bibinfo {title} {Spin polarization of electrons by
  nonmagnetic heterostructures: The basics of spin optics},\ }\href
  {https://doi.org/10.1103/PhysRevLett.92.086602} {\bibfield  {journal}
  {\bibinfo  {journal} {Phys. Rev. Lett.}\ }\textbf {\bibinfo {volume} {92}},\
  \bibinfo {pages} {086602} (\bibinfo {year} {2004})}\BibitemShut {NoStop}%
\bibitem [{\citenamefont {Xiao}\ \emph {et~al.}(2012)\citenamefont {Xiao},
  \citenamefont {Liu}, \citenamefont {Feng}, \citenamefont {Xu},\ and\
  \citenamefont {Yao}}]{nonMag_spinValley_2012_prl}%
  \BibitemOpen
  \bibfield  {author} {\bibinfo {author} {\bibfnamefont {D.}~\bibnamefont
  {Xiao}}, \bibinfo {author} {\bibfnamefont {G.-B.}\ \bibnamefont {Liu}},
  \bibinfo {author} {\bibfnamefont {W.}~\bibnamefont {Feng}}, \bibinfo {author}
  {\bibfnamefont {X.}~\bibnamefont {Xu}},\ and\ \bibinfo {author}
  {\bibfnamefont {W.}~\bibnamefont {Yao}},\ }\bibfield  {title} {\bibinfo
  {title} {Coupled spin and valley physics in monolayers of
  ${\mathrm{mos}}_{2}$ and other group-vi dichalcogenides},\ }\href
  {https://doi.org/10.1103/PhysRevLett.108.196802} {\bibfield  {journal}
  {\bibinfo  {journal} {Phys. Rev. Lett.}\ }\textbf {\bibinfo {volume} {108}},\
  \bibinfo {pages} {196802} (\bibinfo {year} {2012})}\BibitemShut {NoStop}%
\bibitem [{\citenamefont {Zhang}\ \emph {et~al.}(2014)\citenamefont {Zhang},
  \citenamefont {Liu}, \citenamefont {Luo}, \citenamefont {Freeman},\ and\
  \citenamefont {Zunger}}]{hiddenspin_2014_np}%
  \BibitemOpen
  \bibfield  {author} {\bibinfo {author} {\bibfnamefont {X.}~\bibnamefont
  {Zhang}}, \bibinfo {author} {\bibfnamefont {Q.}~\bibnamefont {Liu}}, \bibinfo
  {author} {\bibfnamefont {J.-W.}\ \bibnamefont {Luo}}, \bibinfo {author}
  {\bibfnamefont {A.~J.}\ \bibnamefont {Freeman}},\ and\ \bibinfo {author}
  {\bibfnamefont {A.}~\bibnamefont {Zunger}},\ }\bibfield  {title} {\bibinfo
  {title} {Hidden spin polarization in inversion-symmetric bulk crystals},\
  }\href {https://doi.org/10.1038/nphys2933} {\bibfield  {journal} {\bibinfo
  {journal} {Nature Physics}\ }\textbf {\bibinfo {volume} {10}},\ \bibinfo
  {pages} {387–393} (\bibinfo {year} {2014})}\BibitemShut {NoStop}%
\bibitem [{\citenamefont {\ifmmode~\check{Z}\else \v{Z}\fi{}elezn\'y}\ \emph
  {et~al.}(2017)\citenamefont {\ifmmode~\check{Z}\else \v{Z}\fi{}elezn\'y},
  \citenamefont {Zhang}, \citenamefont {Felser},\ and\ \citenamefont
  {Yan}}]{noncollinear_yan_2017}%
  \BibitemOpen
  \bibfield  {author} {\bibinfo {author} {\bibfnamefont {J.}~\bibnamefont
  {\ifmmode~\check{Z}\else \v{Z}\fi{}elezn\'y}}, \bibinfo {author}
  {\bibfnamefont {Y.}~\bibnamefont {Zhang}}, \bibinfo {author} {\bibfnamefont
  {C.}~\bibnamefont {Felser}},\ and\ \bibinfo {author} {\bibfnamefont
  {B.}~\bibnamefont {Yan}},\ }\bibfield  {title} {\bibinfo {title}
  {Spin-polarized current in noncollinear antiferromagnets},\ }\href
  {https://doi.org/10.1103/PhysRevLett.119.187204} {\bibfield  {journal}
  {\bibinfo  {journal} {Phys. Rev. Lett.}\ }\textbf {\bibinfo {volume} {119}},\
  \bibinfo {pages} {187204} (\bibinfo {year} {2017})}\BibitemShut {NoStop}%
\bibitem [{\citenamefont {{Tian}}\ \emph {et~al.}(2017)\citenamefont {{Tian}},
  \citenamefont {{Hong}}, \citenamefont {{Miotkowski}}, \citenamefont
  {{Datta}},\ and\ \citenamefont {{Chen}}}]{TI_spincurrent_2017}%
  \BibitemOpen
  \bibfield  {author} {\bibinfo {author} {\bibfnamefont {J.}~\bibnamefont
  {{Tian}}}, \bibinfo {author} {\bibfnamefont {S.}~\bibnamefont {{Hong}}},
  \bibinfo {author} {\bibfnamefont {I.}~\bibnamefont {{Miotkowski}}}, \bibinfo
  {author} {\bibfnamefont {S.}~\bibnamefont {{Datta}}},\ and\ \bibinfo {author}
  {\bibfnamefont {Y.~P.}\ \bibnamefont {{Chen}}},\ }\bibfield  {title}
  {\bibinfo {title} {{Observation of current-induced, long-lived persistent
  spin polarization in a topological insulator: A rechargeable spin battery}},\
  }\href {https://doi.org/10.1126/sciadv.1602531} {\bibfield  {journal}
  {\bibinfo  {journal} {Science Advances}\ }\textbf {\bibinfo {volume} {3}},\
  \bibinfo {pages} {e1602531} (\bibinfo {year} {2017})}\BibitemShut {NoStop}%
\bibitem [{\citenamefont {Kumar}\ \emph {et~al.}(2019)\citenamefont {Kumar},
  \citenamefont {Kumawat},\ and\ \citenamefont {Pathak}}]{ferro_nanowire_2019}%
  \BibitemOpen
  \bibfield  {author} {\bibinfo {author} {\bibfnamefont {S.}~\bibnamefont
  {Kumar}}, \bibinfo {author} {\bibfnamefont {R.~L.}\ \bibnamefont {Kumawat}},\
  and\ \bibinfo {author} {\bibfnamefont {B.}~\bibnamefont {Pathak}},\
  }\bibfield  {title} {\bibinfo {title} {Spin-polarized current in
  ferromagnetic half-metallic transition-metal iodide nanowires},\ }\href
  {https://doi.org/10.1021/acs.jpcc.9b02464} {\bibfield  {journal} {\bibinfo
  {journal} {The Journal of Physical Chemistry C}\ }\textbf {\bibinfo {volume}
  {123}},\ \bibinfo {pages} {15717} (\bibinfo {year} {2019})}\BibitemShut
  {NoStop}%
\bibitem [{\citenamefont {Zhang}\ \emph {et~al.}(2020)\citenamefont {Zhang},
  \citenamefont {Liu}, \citenamefont {He}, \citenamefont {Zhang}, \citenamefont
  {Chen}, \citenamefont {Tong}, \citenamefont {Huang}, \citenamefont {Zhou},
  \citenamefont {Zheng}, \citenamefont {Chen}, \citenamefont {Braun},
  \citenamefont {Meixner},\ and\ \citenamefont {Pan}}]{vdw_2020_nc}%
  \BibitemOpen
  \bibfield  {author} {\bibinfo {author} {\bibfnamefont {D.}~\bibnamefont
  {Zhang}}, \bibinfo {author} {\bibfnamefont {Y.}~\bibnamefont {Liu}}, \bibinfo
  {author} {\bibfnamefont {M.}~\bibnamefont {He}}, \bibinfo {author}
  {\bibfnamefont {A.}~\bibnamefont {Zhang}}, \bibinfo {author} {\bibfnamefont
  {S.}~\bibnamefont {Chen}}, \bibinfo {author} {\bibfnamefont {Q.}~\bibnamefont
  {Tong}}, \bibinfo {author} {\bibfnamefont {L.}~\bibnamefont {Huang}},
  \bibinfo {author} {\bibfnamefont {Z.}~\bibnamefont {Zhou}}, \bibinfo {author}
  {\bibfnamefont {W.}~\bibnamefont {Zheng}}, \bibinfo {author} {\bibfnamefont
  {M.}~\bibnamefont {Chen}}, \bibinfo {author} {\bibfnamefont {K.}~\bibnamefont
  {Braun}}, \bibinfo {author} {\bibfnamefont {X.}~\bibnamefont {Meixner},
  \bibfnamefont {A.~amd~Wang}},\ and\ \bibinfo {author} {\bibfnamefont
  {A.}~\bibnamefont {Pan}},\ }\bibfield  {title} {\bibinfo {title} {Room
  temperature near unity spin polarization in 2d van der waals
  heterostructures},\ }\bibfield  {journal} {\bibinfo  {journal} {Nature
  Communications}\ }\textbf {\bibinfo {volume} {11}},\ \href
  {https://doi.org/10.1038/s41467-020-18307-w} {10.1038/s41467-020-18307-w}
  (\bibinfo {year} {2020})\BibitemShut {NoStop}%
\bibitem [{\citenamefont {Nan}\ \emph {et~al.}(2020)\citenamefont {Nan},
  \citenamefont {Quintela}, \citenamefont {Irwin}, \citenamefont {Gurung},
  \citenamefont {Shao}, \citenamefont {Gibbons}, \citenamefont {Campbell},
  \citenamefont {Song}, \citenamefont {Choi}, \citenamefont {Guo},\ and\
  \citenamefont {et~al.}}]{afm_2020_nc}%
  \BibitemOpen
  \bibfield  {author} {\bibinfo {author} {\bibfnamefont {T.}~\bibnamefont
  {Nan}}, \bibinfo {author} {\bibfnamefont {C.~X.}\ \bibnamefont {Quintela}},
  \bibinfo {author} {\bibfnamefont {J.}~\bibnamefont {Irwin}}, \bibinfo
  {author} {\bibfnamefont {G.}~\bibnamefont {Gurung}}, \bibinfo {author}
  {\bibfnamefont {D.~F.}\ \bibnamefont {Shao}}, \bibinfo {author}
  {\bibfnamefont {J.}~\bibnamefont {Gibbons}}, \bibinfo {author} {\bibfnamefont
  {N.}~\bibnamefont {Campbell}}, \bibinfo {author} {\bibfnamefont
  {K.}~\bibnamefont {Song}}, \bibinfo {author} {\bibfnamefont {S.~Y.}\
  \bibnamefont {Choi}}, \bibinfo {author} {\bibfnamefont {L.}~\bibnamefont
  {Guo}},\ and\ \bibinfo {author} {\bibnamefont {et~al.}},\ }\bibfield  {title}
  {\bibinfo {title} {Controlling spin current polarization through
  non-collinear antiferromagnetism},\ }\bibfield  {journal} {\bibinfo
  {journal} {Nature Communications}\ }\textbf {\bibinfo {volume} {11}},\ \href
  {https://doi.org/10.1038/s41467-020-17999-4} {10.1038/s41467-020-17999-4}
  (\bibinfo {year} {2020})\BibitemShut {NoStop}%
\bibitem [{\citenamefont {Yuan}\ \emph {et~al.}(2020)\citenamefont {Yuan},
  \citenamefont {Wang}, \citenamefont {Luo}, \citenamefont {Rashba},\ and\
  \citenamefont {Zunger}}]{noncollinear_2020_prb}%
  \BibitemOpen
  \bibfield  {author} {\bibinfo {author} {\bibfnamefont {L.-D.}\ \bibnamefont
  {Yuan}}, \bibinfo {author} {\bibfnamefont {Z.}~\bibnamefont {Wang}}, \bibinfo
  {author} {\bibfnamefont {J.-W.}\ \bibnamefont {Luo}}, \bibinfo {author}
  {\bibfnamefont {E.~I.}\ \bibnamefont {Rashba}},\ and\ \bibinfo {author}
  {\bibfnamefont {A.}~\bibnamefont {Zunger}},\ }\bibfield  {title} {\bibinfo
  {title} {Giant momentum-dependent spin splitting in centrosymmetric low-$z$
  antiferromagnets},\ }\href {https://doi.org/10.1103/PhysRevB.102.014422}
  {\bibfield  {journal} {\bibinfo  {journal} {Phys. Rev. B}\ }\textbf {\bibinfo
  {volume} {102}},\ \bibinfo {pages} {014422} (\bibinfo {year}
  {2020})}\BibitemShut {NoStop}%
\bibitem [{\citenamefont {Davidson}\ \emph {et~al.}(2020)\citenamefont
  {Davidson}, \citenamefont {Amin}, \citenamefont {Aljuaid}, \citenamefont
  {Haney},\ and\ \citenamefont {Fan}}]{fm_2020_plA}%
  \BibitemOpen
  \bibfield  {author} {\bibinfo {author} {\bibfnamefont {A.}~\bibnamefont
  {Davidson}}, \bibinfo {author} {\bibfnamefont {V.~P.}\ \bibnamefont {Amin}},
  \bibinfo {author} {\bibfnamefont {W.~S.}\ \bibnamefont {Aljuaid}}, \bibinfo
  {author} {\bibfnamefont {P.~M.}\ \bibnamefont {Haney}},\ and\ \bibinfo
  {author} {\bibfnamefont {X.}~\bibnamefont {Fan}},\ }\bibfield  {title}
  {\bibinfo {title} {Perspectives of electrically generated spin currents in
  ferromagnetic materials},\ }\href
  {https://doi.org/10.1016/j.physleta.2019.126228} {\bibfield  {journal}
  {\bibinfo  {journal} {Physics Letters A}\ }\textbf {\bibinfo {volume}
  {384}},\ \bibinfo {pages} {126228} (\bibinfo {year} {2020})}\BibitemShut
  {NoStop}%
\bibitem [{\citenamefont {Gonz\'alez-Hern\'andez}\ \emph
  {et~al.}(2021)\citenamefont {Gonz\'alez-Hern\'andez}, \citenamefont
  {\ifmmode~\check{S}\else \v{S}\fi{}mejkal}, \citenamefont {V\'yborn\'y},
  \citenamefont {Yahagi}, \citenamefont {Sinova}, \citenamefont {Jungwirth},\
  and\ \citenamefont {\ifmmode~\check{Z}\else
  \v{Z}\fi{}elezn\'y}}]{afm_sSplitter_2021_Jario}%
  \BibitemOpen
  \bibfield  {author} {\bibinfo {author} {\bibfnamefont {R.}~\bibnamefont
  {Gonz\'alez-Hern\'andez}}, \bibinfo {author} {\bibfnamefont {L.}~\bibnamefont
  {\ifmmode~\check{S}\else \v{S}\fi{}mejkal}}, \bibinfo {author} {\bibfnamefont
  {K.}~\bibnamefont {V\'yborn\'y}}, \bibinfo {author} {\bibfnamefont
  {Y.}~\bibnamefont {Yahagi}}, \bibinfo {author} {\bibfnamefont
  {J.}~\bibnamefont {Sinova}}, \bibinfo {author} {\bibfnamefont {T.~c.~v.}\
  \bibnamefont {Jungwirth}},\ and\ \bibinfo {author} {\bibfnamefont
  {J.}~\bibnamefont {\ifmmode~\check{Z}\else \v{Z}\fi{}elezn\'y}},\ }\bibfield
  {title} {\bibinfo {title} {Efficient electrical spin splitter based on
  nonrelativistic collinear antiferromagnetism},\ }\href
  {https://doi.org/10.1103/PhysRevLett.126.127701} {\bibfield  {journal}
  {\bibinfo  {journal} {Phys. Rev. Lett.}\ }\textbf {\bibinfo {volume} {126}},\
  \bibinfo {pages} {127701} (\bibinfo {year} {2021})}\BibitemShut {NoStop}%
\bibitem [{\citenamefont {{Huang}}\ \emph {et~al.}(2021)\citenamefont
  {{Huang}}, \citenamefont {{Poloj{\"a}rvi}}, \citenamefont {{Hiura}},
  \citenamefont {{H{\"o}jer}}, \citenamefont {{Aho}}, \citenamefont {{Isoaho}},
  \citenamefont {{Hakkarainen}}, \citenamefont {{Guina}}, \citenamefont
  {{Sato}}, \citenamefont {{Takayama}}, \citenamefont {{Murayama}},
  \citenamefont {{Buyanova}},\ and\ \citenamefont {{Chen}}}]{optospin_2021}%
  \BibitemOpen
  \bibfield  {author} {\bibinfo {author} {\bibfnamefont {Y.}~\bibnamefont
  {{Huang}}}, \bibinfo {author} {\bibfnamefont {V.}~\bibnamefont
  {{Poloj{\"a}rvi}}}, \bibinfo {author} {\bibfnamefont {S.}~\bibnamefont
  {{Hiura}}}, \bibinfo {author} {\bibfnamefont {P.}~\bibnamefont
  {{H{\"o}jer}}}, \bibinfo {author} {\bibfnamefont {A.}~\bibnamefont {{Aho}}},
  \bibinfo {author} {\bibfnamefont {R.}~\bibnamefont {{Isoaho}}}, \bibinfo
  {author} {\bibfnamefont {T.}~\bibnamefont {{Hakkarainen}}}, \bibinfo {author}
  {\bibfnamefont {M.}~\bibnamefont {{Guina}}}, \bibinfo {author} {\bibfnamefont
  {S.}~\bibnamefont {{Sato}}}, \bibinfo {author} {\bibfnamefont
  {J.}~\bibnamefont {{Takayama}}}, \bibinfo {author} {\bibfnamefont
  {A.}~\bibnamefont {{Murayama}}}, \bibinfo {author} {\bibfnamefont {I.~A.}\
  \bibnamefont {{Buyanova}}},\ and\ \bibinfo {author} {\bibfnamefont {W.~M.}\
  \bibnamefont {{Chen}}},\ }\bibfield  {title} {\bibinfo {title}
  {{Room-temperature electron spin polarization exceeding 90\% in an
  opto-spintronic semiconductor nanostructure via remote spin filtering}},\
  }\href {https://doi.org/10.1038/s41566-021-00786-y} {\bibfield  {journal}
  {\bibinfo  {journal} {Nature Photonics}\ }\textbf {\bibinfo {volume} {15}},\
  \bibinfo {pages} {475} (\bibinfo {year} {2021})}\BibitemShut {NoStop}%
\bibitem [{\citenamefont {Xu}\ \emph {et~al.}(2021)\citenamefont {Xu},
  \citenamefont {Wang}, \citenamefont {Zhou},\ and\ \citenamefont
  {Li}}]{spinPVE_2021_nc}%
  \BibitemOpen
  \bibfield  {author} {\bibinfo {author} {\bibfnamefont {H.}~\bibnamefont
  {Xu}}, \bibinfo {author} {\bibfnamefont {H.}~\bibnamefont {Wang}}, \bibinfo
  {author} {\bibfnamefont {J.}~\bibnamefont {Zhou}},\ and\ \bibinfo {author}
  {\bibfnamefont {J.}~\bibnamefont {Li}},\ }\bibfield  {title} {\bibinfo
  {title} {Pure spin photocurrent in non-centrosymmetric crystals: bulk spin
  photovoltaic effect},\ }\bibfield  {journal} {\bibinfo  {journal} {Nature
  Communications}\ }\textbf {\bibinfo {volume} {12}},\ \href
  {https://doi.org/10.1038/s41467-021-24541-7} {10.1038/s41467-021-24541-7}
  (\bibinfo {year} {2021})\BibitemShut {NoStop}%
\bibitem [{\citenamefont {Hayami}\ \emph {et~al.}(2022)\citenamefont {Hayami},
  \citenamefont {Yatsushiro},\ and\ \citenamefont
  {Kusunose}}]{2022_nlshe_Hayami}%
  \BibitemOpen
  \bibfield  {author} {\bibinfo {author} {\bibfnamefont {S.}~\bibnamefont
  {Hayami}}, \bibinfo {author} {\bibfnamefont {M.}~\bibnamefont {Yatsushiro}},\
  and\ \bibinfo {author} {\bibfnamefont {H.}~\bibnamefont {Kusunose}},\
  }\bibfield  {title} {\bibinfo {title} {Nonlinear spin hall effect in
  $\mathcal{PT}$-symmetric collinear magnets},\ }\href
  {https://doi.org/10.1103/PhysRevB.106.024405} {\bibfield  {journal} {\bibinfo
   {journal} {Phys. Rev. B}\ }\textbf {\bibinfo {volume} {106}},\ \bibinfo
  {pages} {024405} (\bibinfo {year} {2022})}\BibitemShut {NoStop}%
\bibitem [{\citenamefont {Fan}\ and\ \citenamefont
  {Wang}(2016)}]{TIs_spintronics_Wang2016}%
  \BibitemOpen
  \bibfield  {author} {\bibinfo {author} {\bibfnamefont {Y.}~\bibnamefont
  {Fan}}\ and\ \bibinfo {author} {\bibfnamefont {K.~L.}\ \bibnamefont {Wang}},\
  }\bibfield  {title} {\bibinfo {title} {Spintronics based on topological
  insulators},\ }\href {https://doi.org/10.1142/S2010324716400014} {\bibfield
  {journal} {\bibinfo  {journal} {SPIN}\ }\textbf {\bibinfo {volume} {06}},\
  \bibinfo {pages} {1640001} (\bibinfo {year} {2016})}\BibitemShut {NoStop}%
\bibitem [{\citenamefont {{Breunig}}\ and\ \citenamefont
  {{Ando}}(2021)}]{TIs_techreview_2021}%
  \BibitemOpen
  \bibfield  {author} {\bibinfo {author} {\bibfnamefont {O.}~\bibnamefont
  {{Breunig}}}\ and\ \bibinfo {author} {\bibfnamefont {Y.}~\bibnamefont
  {{Ando}}},\ }\bibfield  {title} {\bibinfo {title} {{Opportunities in
  topological insulator devices}},\ }\href
  {https://doi.org/10.1038/s42254-021-00402-6} {\bibfield  {journal} {\bibinfo
  {journal} {Nature Reviews Physics}\ }\textbf {\bibinfo {volume} {4}},\
  \bibinfo {pages} {184} (\bibinfo {year} {2021})}\BibitemShut {NoStop}%
\bibitem [{\citenamefont {{Li}}\ \emph {et~al.}(2014)\citenamefont {{Li}},
  \citenamefont {{van `T Erve}}, \citenamefont {{Robinson}}, \citenamefont
  {{Liu}}, \citenamefont {{Li}},\ and\ \citenamefont
  {{Jonker}}}]{sp_detection_2014_Nnanotech}%
  \BibitemOpen
  \bibfield  {author} {\bibinfo {author} {\bibfnamefont {C.~H.}\ \bibnamefont
  {{Li}}}, \bibinfo {author} {\bibfnamefont {O.~M.~J.}\ \bibnamefont {{van `T
  Erve}}}, \bibinfo {author} {\bibfnamefont {J.~T.}\ \bibnamefont
  {{Robinson}}}, \bibinfo {author} {\bibfnamefont {Y.}~\bibnamefont {{Liu}}},
  \bibinfo {author} {\bibfnamefont {L.}~\bibnamefont {{Li}}},\ and\ \bibinfo
  {author} {\bibfnamefont {B.~T.}\ \bibnamefont {{Jonker}}},\ }\bibfield
  {title} {\bibinfo {title} {{Electrical detection of charge-current-induced
  spin polarization due to spin-momentum locking in Bi$_{2}$Se$_{3}$}},\ }\href
  {https://doi.org/10.1038/nnano.2014.16} {\bibfield  {journal} {\bibinfo
  {journal} {Nature Nanotechnology}\ }\textbf {\bibinfo {volume} {9}},\
  \bibinfo {pages} {218} (\bibinfo {year} {2014})}\BibitemShut {NoStop}%
\bibitem [{\citenamefont {{Ando}}\ \emph {et~al.}(2014)\citenamefont {{Ando}},
  \citenamefont {{Hamasaki}}, \citenamefont {{Kurokawa}}, \citenamefont
  {{Ichiba}}, \citenamefont {{Yang}}, \citenamefont {{Novak}}, \citenamefont
  {{Sasaki}}, \citenamefont {{Segawa}}, \citenamefont {{Ando}},\ and\
  \citenamefont {{Shiraishi}}}]{sp_detect_2014_nanolett}%
  \BibitemOpen
  \bibfield  {author} {\bibinfo {author} {\bibfnamefont {Y.}~\bibnamefont
  {{Ando}}}, \bibinfo {author} {\bibfnamefont {T.}~\bibnamefont {{Hamasaki}}},
  \bibinfo {author} {\bibfnamefont {T.}~\bibnamefont {{Kurokawa}}}, \bibinfo
  {author} {\bibfnamefont {K.}~\bibnamefont {{Ichiba}}}, \bibinfo {author}
  {\bibfnamefont {F.}~\bibnamefont {{Yang}}}, \bibinfo {author} {\bibfnamefont
  {M.}~\bibnamefont {{Novak}}}, \bibinfo {author} {\bibfnamefont
  {S.}~\bibnamefont {{Sasaki}}}, \bibinfo {author} {\bibfnamefont
  {K.}~\bibnamefont {{Segawa}}}, \bibinfo {author} {\bibfnamefont
  {Y.}~\bibnamefont {{Ando}}},\ and\ \bibinfo {author} {\bibfnamefont
  {M.}~\bibnamefont {{Shiraishi}}},\ }\bibfield  {title} {\bibinfo {title}
  {{Electrical Detection of the Spin Polarization Due to Charge Flow in the
  Surface State of the Topological Insulator Bi1.5Sb0.5Te1.7Se1.3}},\ }\href
  {https://doi.org/10.1021/nl502546c} {\bibfield  {journal} {\bibinfo
  {journal} {Nano Letters}\ }\textbf {\bibinfo {volume} {14}},\ \bibinfo
  {pages} {6226} (\bibinfo {year} {2014})}\BibitemShut {NoStop}%
\bibitem [{\citenamefont {Tian}\ \emph {et~al.}(2014)\citenamefont {Tian},
  \citenamefont {Childres}, \citenamefont {Cao}, \citenamefont {Shen},
  \citenamefont {Miotkowski},\ and\ \citenamefont
  {Chen}}]{sp_transport_2014_ssc}%
  \BibitemOpen
  \bibfield  {author} {\bibinfo {author} {\bibfnamefont {J.}~\bibnamefont
  {Tian}}, \bibinfo {author} {\bibfnamefont {I.}~\bibnamefont {Childres}},
  \bibinfo {author} {\bibfnamefont {H.}~\bibnamefont {Cao}}, \bibinfo {author}
  {\bibfnamefont {T.}~\bibnamefont {Shen}}, \bibinfo {author} {\bibfnamefont
  {I.}~\bibnamefont {Miotkowski}},\ and\ \bibinfo {author} {\bibfnamefont
  {Y.~P.}\ \bibnamefont {Chen}},\ }\bibfield  {title} {\bibinfo {title}
  {Topological insulator based spin valve devices: Evidence for spin polarized
  transport of spin-momentum-locked topological surface states},\ }\href
  {https://doi.org/https://doi.org/10.1016/j.ssc.2014.04.005} {\bibfield
  {journal} {\bibinfo  {journal} {Solid State Communications}\ }\textbf
  {\bibinfo {volume} {191}},\ \bibinfo {pages} {1} (\bibinfo {year}
  {2014})}\BibitemShut {NoStop}%
\bibitem [{\citenamefont {{Tian}}\ \emph {et~al.}(2015)\citenamefont {{Tian}},
  \citenamefont {{Miotkowski}}, \citenamefont {{Hong}},\ and\ \citenamefont
  {{Chen}}}]{sp_detection_2015_sr}%
  \BibitemOpen
  \bibfield  {author} {\bibinfo {author} {\bibfnamefont {J.}~\bibnamefont
  {{Tian}}}, \bibinfo {author} {\bibfnamefont {I.}~\bibnamefont
  {{Miotkowski}}}, \bibinfo {author} {\bibfnamefont {S.}~\bibnamefont
  {{Hong}}},\ and\ \bibinfo {author} {\bibfnamefont {Y.~P.}\ \bibnamefont
  {{Chen}}},\ }\bibfield  {title} {\bibinfo {title} {{Electrical injection and
  detection of spin-polarized currents in topological insulator
  Bi$_{2}$Te$_{2}$Se}},\ }\href {https://doi.org/10.1038/srep14293} {\bibfield
  {journal} {\bibinfo  {journal} {Scientific Reports}\ }\textbf {\bibinfo
  {volume} {5}},\ \bibinfo {eid} {14293} (\bibinfo {year} {2015})}\BibitemShut
  {NoStop}%
\bibitem [{\citenamefont {{Dankert}}\ \emph {et~al.}(2015)\citenamefont
  {{Dankert}}, \citenamefont {{Geurs}}, \citenamefont {{Kamalakar}},
  \citenamefont {{Charpentier}},\ and\ \citenamefont
  {{Dash}}}]{sp_current_2015_nanolett}%
  \BibitemOpen
  \bibfield  {author} {\bibinfo {author} {\bibfnamefont {A.}~\bibnamefont
  {{Dankert}}}, \bibinfo {author} {\bibfnamefont {J.}~\bibnamefont {{Geurs}}},
  \bibinfo {author} {\bibfnamefont {M.~V.}\ \bibnamefont {{Kamalakar}}},
  \bibinfo {author} {\bibfnamefont {S.}~\bibnamefont {{Charpentier}}},\ and\
  \bibinfo {author} {\bibfnamefont {S.~P.}\ \bibnamefont {{Dash}}},\ }\bibfield
   {title} {\bibinfo {title} {{Room Temperature Electrical Detection of Spin
  Polarized Currents in Topological Insulators}},\ }\href
  {https://doi.org/10.1021/acs.nanolett.5b03080} {\bibfield  {journal}
  {\bibinfo  {journal} {Nano Letters}\ }\textbf {\bibinfo {volume} {15}},\
  \bibinfo {pages} {7976} (\bibinfo {year} {2015})}\BibitemShut {NoStop}%
\bibitem [{\citenamefont {Liu}\ \emph {et~al.}(2015)\citenamefont {Liu},
  \citenamefont {Richardella}, \citenamefont {Garate}, \citenamefont {Zhu},
  \citenamefont {Samarth},\ and\ \citenamefont {Chen}}]{sp_tunneling_2015_prb}%
  \BibitemOpen
  \bibfield  {author} {\bibinfo {author} {\bibfnamefont {L.}~\bibnamefont
  {Liu}}, \bibinfo {author} {\bibfnamefont {A.}~\bibnamefont {Richardella}},
  \bibinfo {author} {\bibfnamefont {I.}~\bibnamefont {Garate}}, \bibinfo
  {author} {\bibfnamefont {Y.}~\bibnamefont {Zhu}}, \bibinfo {author}
  {\bibfnamefont {N.}~\bibnamefont {Samarth}},\ and\ \bibinfo {author}
  {\bibfnamefont {C.-T.}\ \bibnamefont {Chen}},\ }\bibfield  {title} {\bibinfo
  {title} {Spin-polarized tunneling study of spin-momentum locking in
  topological insulators},\ }\href {https://doi.org/10.1103/PhysRevB.91.235437}
  {\bibfield  {journal} {\bibinfo  {journal} {Phys. Rev. B}\ }\textbf {\bibinfo
  {volume} {91}},\ \bibinfo {pages} {235437} (\bibinfo {year}
  {2015})}\BibitemShut {NoStop}%
\bibitem [{\citenamefont {Lee}\ \emph {et~al.}(2015)\citenamefont {Lee},
  \citenamefont {Richardella}, \citenamefont {Hickey}, \citenamefont
  {Mkhoyan},\ and\ \citenamefont {Samarth}}]{sp_mapping_2015_prb}%
  \BibitemOpen
  \bibfield  {author} {\bibinfo {author} {\bibfnamefont {J.~S.}\ \bibnamefont
  {Lee}}, \bibinfo {author} {\bibfnamefont {A.}~\bibnamefont {Richardella}},
  \bibinfo {author} {\bibfnamefont {D.~R.}\ \bibnamefont {Hickey}}, \bibinfo
  {author} {\bibfnamefont {K.~A.}\ \bibnamefont {Mkhoyan}},\ and\ \bibinfo
  {author} {\bibfnamefont {N.}~\bibnamefont {Samarth}},\ }\bibfield  {title}
  {\bibinfo {title} {Mapping the chemical potential dependence of
  current-induced spin polarization in a topological insulator},\ }\href
  {https://doi.org/10.1103/PhysRevB.92.155312} {\bibfield  {journal} {\bibinfo
  {journal} {Phys. Rev. B}\ }\textbf {\bibinfo {volume} {92}},\ \bibinfo
  {pages} {155312} (\bibinfo {year} {2015})}\BibitemShut {NoStop}%
\bibitem [{\citenamefont {{Li}}\ \emph {et~al.}(2016)\citenamefont {{Li}},
  \citenamefont {{van `T Erve}}, \citenamefont {{Rajput}}, \citenamefont
  {{Li}},\ and\ \citenamefont {{Jonker}}}]{sp_comparison_2016_nc}%
  \BibitemOpen
  \bibfield  {author} {\bibinfo {author} {\bibfnamefont {C.~H.}\ \bibnamefont
  {{Li}}}, \bibinfo {author} {\bibfnamefont {O.~M.~J.}\ \bibnamefont {{van `T
  Erve}}}, \bibinfo {author} {\bibfnamefont {S.}~\bibnamefont {{Rajput}}},
  \bibinfo {author} {\bibfnamefont {L.}~\bibnamefont {{Li}}},\ and\ \bibinfo
  {author} {\bibfnamefont {B.~T.}\ \bibnamefont {{Jonker}}},\ }\bibfield
  {title} {\bibinfo {title} {{Direct comparison of current-induced spin
  polarization in topological insulator Bi$_{2}$Se$_{3}$ and InAs Rashba
  states}},\ }\href {https://doi.org/10.1038/ncomms13518} {\bibfield  {journal}
  {\bibinfo  {journal} {Nature Communications}\ }\textbf {\bibinfo {volume}
  {7}},\ \bibinfo {eid} {13518} (\bibinfo {year} {2016})}\BibitemShut {NoStop}%
\bibitem [{\citenamefont {{Maa{\ss}}}\ \emph {et~al.}(2016)\citenamefont
  {{Maa{\ss}}}, \citenamefont {{Bentmann}}, \citenamefont {{Seibel}},
  \citenamefont {{Tusche}}, \citenamefont {{Eremeev}}, \citenamefont
  {{Peixoto}}, \citenamefont {{Tereshchenko}}, \citenamefont {{Kokh}},
  \citenamefont {{Chulkov}}, \citenamefont {{Kirschner}},\ and\ \citenamefont
  {{Reinert}}}]{nonMag_sp_2016_nc}%
  \BibitemOpen
  \bibfield  {author} {\bibinfo {author} {\bibfnamefont {H.}~\bibnamefont
  {{Maa{\ss}}}}, \bibinfo {author} {\bibfnamefont {H.}~\bibnamefont
  {{Bentmann}}}, \bibinfo {author} {\bibfnamefont {C.}~\bibnamefont
  {{Seibel}}}, \bibinfo {author} {\bibfnamefont {C.}~\bibnamefont {{Tusche}}},
  \bibinfo {author} {\bibfnamefont {S.~V.}\ \bibnamefont {{Eremeev}}}, \bibinfo
  {author} {\bibfnamefont {T.~R.~F.}\ \bibnamefont {{Peixoto}}}, \bibinfo
  {author} {\bibfnamefont {O.~E.}\ \bibnamefont {{Tereshchenko}}}, \bibinfo
  {author} {\bibfnamefont {K.~A.}\ \bibnamefont {{Kokh}}}, \bibinfo {author}
  {\bibfnamefont {E.~V.}\ \bibnamefont {{Chulkov}}}, \bibinfo {author}
  {\bibfnamefont {J.}~\bibnamefont {{Kirschner}}},\ and\ \bibinfo {author}
  {\bibfnamefont {F.}~\bibnamefont {{Reinert}}},\ }\bibfield  {title} {\bibinfo
  {title} {{Spin-texture inversion in the giant Rashba semiconductor BiTeI}},\
  }\href {https://doi.org/10.1038/ncomms11621} {\bibfield  {journal} {\bibinfo
  {journal} {Nature Communications}\ }\textbf {\bibinfo {volume} {7}},\
  \bibinfo {eid} {11621} (\bibinfo {year} {2016})}\BibitemShut {NoStop}%
\bibitem [{\citenamefont {King}\ \emph {et~al.}(2011)\citenamefont {King},
  \citenamefont {Hatch}, \citenamefont {Bianchi}, \citenamefont {Ovsyannikov},
  \citenamefont {Lupulescu}, \citenamefont {Landolt}, \citenamefont {Slomski},
  \citenamefont {Dil}, \citenamefont {Guan}, \citenamefont {Mi}, \citenamefont
  {Rienks}, \citenamefont {Fink}, \citenamefont {Lindblad}, \citenamefont
  {Svensson}, \citenamefont {Bao}, \citenamefont {Balakrishnan}, \citenamefont
  {Iversen}, \citenamefont {Osterwalder}, \citenamefont {Eberhardt},
  \citenamefont {Baumberger},\ and\ \citenamefont
  {Hofmann}}]{2dEG_2011prl_king}%
  \BibitemOpen
  \bibfield  {author} {\bibinfo {author} {\bibfnamefont {P.~D.~C.}\
  \bibnamefont {King}}, \bibinfo {author} {\bibfnamefont {R.~C.}\ \bibnamefont
  {Hatch}}, \bibinfo {author} {\bibfnamefont {M.}~\bibnamefont {Bianchi}},
  \bibinfo {author} {\bibfnamefont {R.}~\bibnamefont {Ovsyannikov}}, \bibinfo
  {author} {\bibfnamefont {C.}~\bibnamefont {Lupulescu}}, \bibinfo {author}
  {\bibfnamefont {G.}~\bibnamefont {Landolt}}, \bibinfo {author} {\bibfnamefont
  {B.}~\bibnamefont {Slomski}}, \bibinfo {author} {\bibfnamefont {J.~H.}\
  \bibnamefont {Dil}}, \bibinfo {author} {\bibfnamefont {D.}~\bibnamefont
  {Guan}}, \bibinfo {author} {\bibfnamefont {J.~L.}\ \bibnamefont {Mi}},
  \bibinfo {author} {\bibfnamefont {E.~D.~L.}\ \bibnamefont {Rienks}}, \bibinfo
  {author} {\bibfnamefont {J.}~\bibnamefont {Fink}}, \bibinfo {author}
  {\bibfnamefont {A.}~\bibnamefont {Lindblad}}, \bibinfo {author}
  {\bibfnamefont {S.}~\bibnamefont {Svensson}}, \bibinfo {author}
  {\bibfnamefont {S.}~\bibnamefont {Bao}}, \bibinfo {author} {\bibfnamefont
  {G.}~\bibnamefont {Balakrishnan}}, \bibinfo {author} {\bibfnamefont {B.~B.}\
  \bibnamefont {Iversen}}, \bibinfo {author} {\bibfnamefont {J.}~\bibnamefont
  {Osterwalder}}, \bibinfo {author} {\bibfnamefont {W.}~\bibnamefont
  {Eberhardt}}, \bibinfo {author} {\bibfnamefont {F.}~\bibnamefont
  {Baumberger}},\ and\ \bibinfo {author} {\bibfnamefont {P.}~\bibnamefont
  {Hofmann}},\ }\bibfield  {title} {\bibinfo {title} {Large tunable rashba spin
  splitting of a two-dimensional electron gas in
  ${\mathrm{bi}}_{2}{\mathrm{se}}_{3}$},\ }\href
  {https://doi.org/10.1103/PhysRevLett.107.096802} {\bibfield  {journal}
  {\bibinfo  {journal} {Phys. Rev. Lett.}\ }\textbf {\bibinfo {volume} {107}},\
  \bibinfo {pages} {096802} (\bibinfo {year} {2011})}\BibitemShut {NoStop}%
\bibitem [{\citenamefont {Bahramy}\ \emph {et~al.}(2012)\citenamefont
  {Bahramy}, \citenamefont {King}, \citenamefont {De~La~Torre}, \citenamefont
  {Chang}, \citenamefont {Shi}, \citenamefont {Patthey}, \citenamefont
  {Balakrishnan}, \citenamefont {Hofmann}, \citenamefont {Arita}, \citenamefont
  {Nagaosa} \emph {et~al.}}]{2dEG_2012nc_Bahramy}%
  \BibitemOpen
  \bibfield  {author} {\bibinfo {author} {\bibfnamefont {M.}~\bibnamefont
  {Bahramy}}, \bibinfo {author} {\bibfnamefont {P.}~\bibnamefont {King}},
  \bibinfo {author} {\bibfnamefont {A.}~\bibnamefont {De~La~Torre}}, \bibinfo
  {author} {\bibfnamefont {J.}~\bibnamefont {Chang}}, \bibinfo {author}
  {\bibfnamefont {M.}~\bibnamefont {Shi}}, \bibinfo {author} {\bibfnamefont
  {L.}~\bibnamefont {Patthey}}, \bibinfo {author} {\bibfnamefont
  {G.}~\bibnamefont {Balakrishnan}}, \bibinfo {author} {\bibfnamefont
  {P.}~\bibnamefont {Hofmann}}, \bibinfo {author} {\bibfnamefont
  {R.}~\bibnamefont {Arita}}, \bibinfo {author} {\bibfnamefont
  {N.}~\bibnamefont {Nagaosa}}, \emph {et~al.},\ }\bibfield  {title} {\bibinfo
  {title} {Emergent quantum confinement at topological insulator surfaces},\
  }\href@noop {} {\bibfield  {journal} {\bibinfo  {journal} {Nature
  communications}\ }\textbf {\bibinfo {volume} {3}},\ \bibinfo {pages} {1}
  (\bibinfo {year} {2012})}\BibitemShut {NoStop}%
\bibitem [{\citenamefont {de~Vries}\ \emph {et~al.}(2015)\citenamefont
  {de~Vries}, \citenamefont {Kamerbeek}, \citenamefont {Koirala}, \citenamefont
  {Brahlek}, \citenamefont {Salehi}, \citenamefont {Oh}, \citenamefont {van
  Wees},\ and\ \citenamefont {Banerjee}}]{origin_ciSP_2015prb_Banerjee}%
  \BibitemOpen
  \bibfield  {author} {\bibinfo {author} {\bibfnamefont {E.~K.}\ \bibnamefont
  {de~Vries}}, \bibinfo {author} {\bibfnamefont {A.~M.}\ \bibnamefont
  {Kamerbeek}}, \bibinfo {author} {\bibfnamefont {N.}~\bibnamefont {Koirala}},
  \bibinfo {author} {\bibfnamefont {M.}~\bibnamefont {Brahlek}}, \bibinfo
  {author} {\bibfnamefont {M.}~\bibnamefont {Salehi}}, \bibinfo {author}
  {\bibfnamefont {S.}~\bibnamefont {Oh}}, \bibinfo {author} {\bibfnamefont
  {B.~J.}\ \bibnamefont {van Wees}},\ and\ \bibinfo {author} {\bibfnamefont
  {T.}~\bibnamefont {Banerjee}},\ }\bibfield  {title} {\bibinfo {title}
  {Towards the understanding of the origin of charge-current-induced spin
  voltage signals in the topological insulator
  ${\mathrm{bi}}_{2}{\mathrm{se}}_{3}$},\ }\href
  {https://doi.org/10.1103/PhysRevB.92.201102} {\bibfield  {journal} {\bibinfo
  {journal} {Phys. Rev. B}\ }\textbf {\bibinfo {volume} {92}},\ \bibinfo
  {pages} {201102} (\bibinfo {year} {2015})}\BibitemShut {NoStop}%
\bibitem [{\citenamefont {Tian}\ \emph {et~al.}(2021)\citenamefont {Tian},
  \citenamefont {\ifmmode~\mbox{\c{S}}\else \c{S}\fi{}ahin}, \citenamefont
  {Miotkowski}, \citenamefont {Flatt\'e},\ and\ \citenamefont
  {Chen}}]{othermechanism_2021prb_Tian}%
  \BibitemOpen
  \bibfield  {author} {\bibinfo {author} {\bibfnamefont {J.}~\bibnamefont
  {Tian}}, \bibinfo {author} {\bibfnamefont {C.}~\bibnamefont
  {\ifmmode~\mbox{\c{S}}\else \c{S}\fi{}ahin}}, \bibinfo {author}
  {\bibfnamefont {I.}~\bibnamefont {Miotkowski}}, \bibinfo {author}
  {\bibfnamefont {M.~E.}\ \bibnamefont {Flatt\'e}},\ and\ \bibinfo {author}
  {\bibfnamefont {Y.~P.}\ \bibnamefont {Chen}},\ }\bibfield  {title} {\bibinfo
  {title} {Opposite current-induced spin polarizations in bulk-metallic
  ${\mathrm{bi}}_{2}{\mathrm{se}}_{3}$ and bulk-insulating
  ${\mathrm{bi}}_{2}{\mathrm{te}}_{2}\mathrm{Se}$ topological insulator thin
  flakes},\ }\href {https://doi.org/10.1103/PhysRevB.103.035412} {\bibfield
  {journal} {\bibinfo  {journal} {Phys. Rev. B}\ }\textbf {\bibinfo {volume}
  {103}},\ \bibinfo {pages} {035412} (\bibinfo {year} {2021})}\BibitemShut
  {NoStop}%
\bibitem [{\citenamefont {Yu}\ \emph {et~al.}(2021{\natexlab{a}})\citenamefont
  {Yu}, \citenamefont {Zhuang}, \citenamefont {Zhu}, \citenamefont {Chen},
  \citenamefont {Liu}, \citenamefont {Zhang}, \citenamefont {Yin},
  \citenamefont {Cheng}, \citenamefont {Lai}, \citenamefont {He},\ and\
  \citenamefont {Xue}}]{ciSP_2021prb_xueqikun}%
  \BibitemOpen
  \bibfield  {author} {\bibinfo {author} {\bibfnamefont {J.}~\bibnamefont
  {Yu}}, \bibinfo {author} {\bibfnamefont {H.}~\bibnamefont {Zhuang}}, \bibinfo
  {author} {\bibfnamefont {K.}~\bibnamefont {Zhu}}, \bibinfo {author}
  {\bibfnamefont {Y.}~\bibnamefont {Chen}}, \bibinfo {author} {\bibfnamefont
  {Y.}~\bibnamefont {Liu}}, \bibinfo {author} {\bibfnamefont {Y.}~\bibnamefont
  {Zhang}}, \bibinfo {author} {\bibfnamefont {C.}~\bibnamefont {Yin}}, \bibinfo
  {author} {\bibfnamefont {S.}~\bibnamefont {Cheng}}, \bibinfo {author}
  {\bibfnamefont {Y.}~\bibnamefont {Lai}}, \bibinfo {author} {\bibfnamefont
  {K.}~\bibnamefont {He}},\ and\ \bibinfo {author} {\bibfnamefont
  {Q.}~\bibnamefont {Xue}},\ }\bibfield  {title} {\bibinfo {title} {Observation
  of current-induced spin polarization in the topological insulator
  ${\mathrm{bi}}_{2}{\mathrm{te}}_{3}$ via circularly polarized photoconductive
  differential current},\ }\href {https://doi.org/10.1103/PhysRevB.104.045428}
  {\bibfield  {journal} {\bibinfo  {journal} {Phys. Rev. B}\ }\textbf {\bibinfo
  {volume} {104}},\ \bibinfo {pages} {045428} (\bibinfo {year}
  {2021}{\natexlab{a}})}\BibitemShut {NoStop}%
\bibitem [{\citenamefont {Li}\ and\ \citenamefont
  {Appelbaum}(2016)}]{SPinterpreting_2016prb_Li}%
  \BibitemOpen
  \bibfield  {author} {\bibinfo {author} {\bibfnamefont {P.}~\bibnamefont
  {Li}}\ and\ \bibinfo {author} {\bibfnamefont {I.}~\bibnamefont {Appelbaum}},\
  }\bibfield  {title} {\bibinfo {title} {Interpreting current-induced spin
  polarization in topological insulator surface states},\ }\href
  {https://doi.org/10.1103/PhysRevB.93.220404} {\bibfield  {journal} {\bibinfo
  {journal} {Phys. Rev. B}\ }\textbf {\bibinfo {volume} {93}},\ \bibinfo
  {pages} {220404} (\bibinfo {year} {2016})}\BibitemShut {NoStop}%
\bibitem [{\citenamefont {Du}\ \emph {et~al.}(2021)\citenamefont {Du},
  \citenamefont {Lu},\ and\ \citenamefont {Xie}}]{nlPerspective_2021_Xie}%
  \BibitemOpen
  \bibfield  {author} {\bibinfo {author} {\bibfnamefont {Z.~Z.}\ \bibnamefont
  {Du}}, \bibinfo {author} {\bibfnamefont {H.-Z.}\ \bibnamefont {Lu}},\ and\
  \bibinfo {author} {\bibfnamefont {X.~C.}\ \bibnamefont {Xie}},\ }\href@noop
  {} {\bibinfo {title} {Perspective: Nonlinear hall effects}} (\bibinfo {year}
  {2021}),\ \Eprint {https://arxiv.org/abs/2105.10940} {arXiv:2105.10940
  [cond-mat.mes-hall]} \BibitemShut {NoStop}%
\bibitem [{\citenamefont {Zeng}\ \emph {et~al.}(2019)\citenamefont {Zeng},
  \citenamefont {Nandy}, \citenamefont {Taraphder},\ and\ \citenamefont
  {Tewari}}]{nlNernst_2019_zeng}%
  \BibitemOpen
  \bibfield  {author} {\bibinfo {author} {\bibfnamefont {C.}~\bibnamefont
  {Zeng}}, \bibinfo {author} {\bibfnamefont {S.}~\bibnamefont {Nandy}},
  \bibinfo {author} {\bibfnamefont {A.}~\bibnamefont {Taraphder}},\ and\
  \bibinfo {author} {\bibfnamefont {S.}~\bibnamefont {Tewari}},\ }\bibfield
  {title} {\bibinfo {title} {Nonlinear nernst effect in bilayer wte$_{2}$},\
  }\href {https://doi.org/10.1103/PhysRevB.100.245102} {\bibfield  {journal}
  {\bibinfo  {journal} {Phys. Rev. B}\ }\textbf {\bibinfo {volume} {100}},\
  \bibinfo {pages} {245102} (\bibinfo {year} {2019})}\BibitemShut {NoStop}%
\bibitem [{\citenamefont {Yu}\ \emph {et~al.}(2019)\citenamefont {Yu},
  \citenamefont {Zhu}, \citenamefont {You}, \citenamefont {Low},\ and\
  \citenamefont {Su}}]{nlNernst_2019_xqy}%
  \BibitemOpen
  \bibfield  {author} {\bibinfo {author} {\bibfnamefont {X.-Q.}\ \bibnamefont
  {Yu}}, \bibinfo {author} {\bibfnamefont {Z.-G.}\ \bibnamefont {Zhu}},
  \bibinfo {author} {\bibfnamefont {J.-S.}\ \bibnamefont {You}}, \bibinfo
  {author} {\bibfnamefont {T.}~\bibnamefont {Low}},\ and\ \bibinfo {author}
  {\bibfnamefont {G.}~\bibnamefont {Su}},\ }\bibfield  {title} {\bibinfo
  {title} {Topological nonlinear anomalous nernst effect in strained transition
  metal dichalcogenides},\ }\href {https://doi.org/10.1103/PhysRevB.99.201410}
  {\bibfield  {journal} {\bibinfo  {journal} {Phys. Rev. B}\ }\textbf {\bibinfo
  {volume} {99}},\ \bibinfo {pages} {201410} (\bibinfo {year}
  {2019})}\BibitemShut {NoStop}%
\bibitem [{\citenamefont {Zeng}\ \emph {et~al.}(2021)\citenamefont {Zeng},
  \citenamefont {Nandy},\ and\ \citenamefont {Tewari}}]{BCD_Weyl_2021_Zeng}%
  \BibitemOpen
  \bibfield  {author} {\bibinfo {author} {\bibfnamefont {C.}~\bibnamefont
  {Zeng}}, \bibinfo {author} {\bibfnamefont {S.}~\bibnamefont {Nandy}},\ and\
  \bibinfo {author} {\bibfnamefont {S.}~\bibnamefont {Tewari}},\ }\bibfield
  {title} {\bibinfo {title} {Nonlinear transport in weyl semimetals induced by
  berry curvature dipole},\ }\href
  {https://doi.org/10.1103/PhysRevB.103.245119} {\bibfield  {journal} {\bibinfo
   {journal} {Phys. Rev. B}\ }\textbf {\bibinfo {volume} {103}},\ \bibinfo
  {pages} {245119} (\bibinfo {year} {2021})}\BibitemShut {NoStop}%
\bibitem [{\citenamefont {Mandal}\ \emph {et~al.}(2020)\citenamefont {Mandal},
  \citenamefont {Das},\ and\ \citenamefont {Agarwal}}]{nlNernst_magnus_2020}%
  \BibitemOpen
  \bibfield  {author} {\bibinfo {author} {\bibfnamefont {D.}~\bibnamefont
  {Mandal}}, \bibinfo {author} {\bibfnamefont {K.}~\bibnamefont {Das}},\ and\
  \bibinfo {author} {\bibfnamefont {A.}~\bibnamefont {Agarwal}},\ }\bibfield
  {title} {\bibinfo {title} {Magnus nernst and thermal hall effect},\ }\href
  {https://doi.org/10.1103/PhysRevB.102.205414} {\bibfield  {journal} {\bibinfo
   {journal} {Phys. Rev. B}\ }\textbf {\bibinfo {volume} {102}},\ \bibinfo
  {pages} {205414} (\bibinfo {year} {2020})}\BibitemShut {NoStop}%
\bibitem [{\citenamefont {Kondo}\ and\ \citenamefont
  {Akagi}(2021)}]{nl_spinNernst_2021akagi}%
  \BibitemOpen
  \bibfield  {author} {\bibinfo {author} {\bibfnamefont {H.}~\bibnamefont
  {Kondo}}\ and\ \bibinfo {author} {\bibfnamefont {Y.}~\bibnamefont {Akagi}},\
  }\bibfield  {title} {\bibinfo {title} {Nonlinear magnon spin nernst effect in
  antiferromagnets and strain-tunable pure spin current},\ }\bibfield
  {journal} {\bibinfo  {journal} {arXiv}\ }\href
  {https://doi.org/10.48550/arXiv.2109.09464} {10.48550/arXiv.2109.09464}
  (\bibinfo {year} {2021})\BibitemShut {NoStop}%
\bibitem [{\citenamefont {Zeng}\ \emph {et~al.}(2022)\citenamefont {Zeng},
  \citenamefont {Nandy},\ and\ \citenamefont {Tewari}}]{chiral_weyl_2020_zeng}%
  \BibitemOpen
  \bibfield  {author} {\bibinfo {author} {\bibfnamefont {C.}~\bibnamefont
  {Zeng}}, \bibinfo {author} {\bibfnamefont {S.}~\bibnamefont {Nandy}},\ and\
  \bibinfo {author} {\bibfnamefont {S.}~\bibnamefont {Tewari}},\ }\bibfield
  {title} {\bibinfo {title} {Chiral anomaly induced nonlinear nernst and
  thermal hall effects in weyl semimetals},\ }\href
  {https://doi.org/10.1103/PhysRevB.105.125131} {\bibfield  {journal} {\bibinfo
   {journal} {Phys. Rev. B}\ }\textbf {\bibinfo {volume} {105}},\ \bibinfo
  {pages} {125131} (\bibinfo {year} {2022})}\BibitemShut {NoStop}%
\bibitem [{\citenamefont {Yu}\ \emph {et~al.}(2021{\natexlab{b}})\citenamefont
  {Yu}, \citenamefont {Zhu},\ and\ \citenamefont {Su}}]{Yu_nlpNernst_2021}%
  \BibitemOpen
  \bibfield  {author} {\bibinfo {author} {\bibfnamefont {X.-Q.}\ \bibnamefont
  {Yu}}, \bibinfo {author} {\bibfnamefont {Z.-G.}\ \bibnamefont {Zhu}},\ and\
  \bibinfo {author} {\bibfnamefont {G.}~\bibnamefont {Su}},\ }\bibfield
  {title} {\bibinfo {title} {Hexagonal warping induced nonlinear planar nernst
  effect in nonmagnetic topological insulators},\ }\href
  {https://doi.org/10.1103/PhysRevB.103.035410} {\bibfield  {journal} {\bibinfo
   {journal} {Phys. Rev. B}\ }\textbf {\bibinfo {volume} {103}},\ \bibinfo
  {pages} {035410} (\bibinfo {year} {2021}{\natexlab{b}})}\BibitemShut
  {NoStop}%
\bibitem [{\citenamefont {He}\ \emph {et~al.}(2019)\citenamefont {He},
  \citenamefont {Zhang}, \citenamefont {Zhu}, \citenamefont {Shi},
  \citenamefont {Heinonen}, \citenamefont {Vignale},\ and\ \citenamefont
  {Yang}}]{NLPHE_2019_prl}%
  \BibitemOpen
  \bibfield  {author} {\bibinfo {author} {\bibfnamefont {P.}~\bibnamefont
  {He}}, \bibinfo {author} {\bibfnamefont {S.~S.-L.}\ \bibnamefont {Zhang}},
  \bibinfo {author} {\bibfnamefont {D.}~\bibnamefont {Zhu}}, \bibinfo {author}
  {\bibfnamefont {S.}~\bibnamefont {Shi}}, \bibinfo {author} {\bibfnamefont
  {O.~G.}\ \bibnamefont {Heinonen}}, \bibinfo {author} {\bibfnamefont
  {G.}~\bibnamefont {Vignale}},\ and\ \bibinfo {author} {\bibfnamefont
  {H.}~\bibnamefont {Yang}},\ }\bibfield  {title} {\bibinfo {title} {Nonlinear
  planar hall effect},\ }\href {https://doi.org/10.1103/PhysRevLett.123.016801}
  {\bibfield  {journal} {\bibinfo  {journal} {Phys. Rev. Lett.}\ }\textbf
  {\bibinfo {volume} {123}},\ \bibinfo {pages} {016801} (\bibinfo {year}
  {2019})}\BibitemShut {NoStop}%
\bibitem [{\citenamefont {Fu}(2009)}]{Liang_2009_warping}%
  \BibitemOpen
  \bibfield  {author} {\bibinfo {author} {\bibfnamefont {L.}~\bibnamefont
  {Fu}},\ }\bibfield  {title} {\bibinfo {title} {Hexagonal warping effects in
  the surface states of the topological insulator
  ${\mathrm{bi}}_{2}{\mathrm{te}}_{3}$},\ }\href
  {https://doi.org/10.1103/PhysRevLett.103.266801} {\bibfield  {journal}
  {\bibinfo  {journal} {Phys. Rev. Lett.}\ }\textbf {\bibinfo {volume} {103}},\
  \bibinfo {pages} {266801} (\bibinfo {year} {2009})}\BibitemShut {NoStop}%
\bibitem [{\citenamefont {Zhang}\ \emph {et~al.}(2012)\citenamefont {Zhang},
  \citenamefont {Kane},\ and\ \citenamefont {Mele}}]{tilt_TSS_2012_zhang}%
  \BibitemOpen
  \bibfield  {author} {\bibinfo {author} {\bibfnamefont {F.}~\bibnamefont
  {Zhang}}, \bibinfo {author} {\bibfnamefont {C.~L.}\ \bibnamefont {Kane}},\
  and\ \bibinfo {author} {\bibfnamefont {E.~J.}\ \bibnamefont {Mele}},\
  }\bibfield  {title} {\bibinfo {title} {Surface states of topological
  insulators},\ }\href {https://doi.org/10.1103/PhysRevB.86.081303} {\bibfield
  {journal} {\bibinfo  {journal} {Phys. Rev. B}\ }\textbf {\bibinfo {volume}
  {86}},\ \bibinfo {pages} {081303} (\bibinfo {year} {2012})}\BibitemShut
  {NoStop}%
\bibitem [{\citenamefont {Zhang}\ \emph {et~al.}(2013)\citenamefont {Zhang},
  \citenamefont {Kane},\ and\ \citenamefont
  {Mele}}]{breakingTRS_Zhang_2013PRL}%
  \BibitemOpen
  \bibfield  {author} {\bibinfo {author} {\bibfnamefont {F.}~\bibnamefont
  {Zhang}}, \bibinfo {author} {\bibfnamefont {C.~L.}\ \bibnamefont {Kane}},\
  and\ \bibinfo {author} {\bibfnamefont {E.~J.}\ \bibnamefont {Mele}},\
  }\bibfield  {title} {\bibinfo {title} {Surface state magnetization and chiral
  edge states on topological insulators},\ }\href
  {https://doi.org/10.1103/PhysRevLett.110.046404} {\bibfield  {journal}
  {\bibinfo  {journal} {Phys. Rev. Lett.}\ }\textbf {\bibinfo {volume} {110}},\
  \bibinfo {pages} {046404} (\bibinfo {year} {2013})}\BibitemShut {NoStop}%
\bibitem [{\citenamefont {He}\ \emph {et~al.}(2018)\citenamefont {He},
  \citenamefont {Zhang}, \citenamefont {Zhu}, \citenamefont {Liu},
  \citenamefont {Wang}, \citenamefont {Yu}, \citenamefont {Vignale},\ and\
  \citenamefont {Yang}}]{bimer_2018_np}%
  \BibitemOpen
  \bibfield  {author} {\bibinfo {author} {\bibfnamefont {P.}~\bibnamefont
  {He}}, \bibinfo {author} {\bibfnamefont {S.~S.-L.}\ \bibnamefont {Zhang}},
  \bibinfo {author} {\bibfnamefont {D.}~\bibnamefont {Zhu}}, \bibinfo {author}
  {\bibfnamefont {Y.}~\bibnamefont {Liu}}, \bibinfo {author} {\bibfnamefont
  {Y.}~\bibnamefont {Wang}}, \bibinfo {author} {\bibfnamefont {J.}~\bibnamefont
  {Yu}}, \bibinfo {author} {\bibfnamefont {G.}~\bibnamefont {Vignale}},\ and\
  \bibinfo {author} {\bibfnamefont {H.}~\bibnamefont {Yang}},\ }\bibfield
  {title} {\bibinfo {title} {Bilinear magnetoelectric resistance as a probe of
  three-dimensional spin texture in topological surface states},\ }\href
  {https://doi.org/10.1038/s41567-017-0039-y} {\bibfield  {journal} {\bibinfo
  {journal} {Nature Physics}\ }\textbf {\bibinfo {volume} {14}},\ \bibinfo
  {pages} {495–499} (\bibinfo {year} {2018})}\BibitemShut {NoStop}%
\bibitem [{\citenamefont {Hamamoto}\ \emph {et~al.}(2017)\citenamefont
  {Hamamoto}, \citenamefont {Ezawa}, \citenamefont {Kim}, \citenamefont
  {Morimoto},\ and\ \citenamefont {Nagaosa}}]{Nagaosa_spin_2017}%
  \BibitemOpen
  \bibfield  {author} {\bibinfo {author} {\bibfnamefont {K.}~\bibnamefont
  {Hamamoto}}, \bibinfo {author} {\bibfnamefont {M.}~\bibnamefont {Ezawa}},
  \bibinfo {author} {\bibfnamefont {K.~W.}\ \bibnamefont {Kim}}, \bibinfo
  {author} {\bibfnamefont {T.}~\bibnamefont {Morimoto}},\ and\ \bibinfo
  {author} {\bibfnamefont {N.}~\bibnamefont {Nagaosa}},\ }\bibfield  {title}
  {\bibinfo {title} {Nonlinear spin current generation in noncentrosymmetric
  spin-orbit coupled systems},\ }\href
  {https://doi.org/10.1103/PhysRevB.95.224430} {\bibfield  {journal} {\bibinfo
  {journal} {Phys. Rev. B}\ }\textbf {\bibinfo {volume} {95}},\ \bibinfo
  {pages} {224430} (\bibinfo {year} {2017})}\BibitemShut {NoStop}%
\bibitem [{\citenamefont {Pan}\ and\ \citenamefont
  {Marinescu}(2019)}]{Ani_spin_2019}%
  \BibitemOpen
  \bibfield  {author} {\bibinfo {author} {\bibfnamefont {A.}~\bibnamefont
  {Pan}}\ and\ \bibinfo {author} {\bibfnamefont {D.~C.}\ \bibnamefont
  {Marinescu}},\ }\bibfield  {title} {\bibinfo {title} {Nonlinear spin-current
  generation in quantum wells with arbitrary rashba-dresselhaus spin-orbit
  interactions},\ }\href {https://doi.org/10.1103/PhysRevB.99.245204}
  {\bibfield  {journal} {\bibinfo  {journal} {Phys. Rev. B}\ }\textbf {\bibinfo
  {volume} {99}},\ \bibinfo {pages} {245204} (\bibinfo {year}
  {2019})}\BibitemShut {NoStop}%
\bibitem [{\citenamefont {Zheng}\ \emph {et~al.}(2020)\citenamefont {Zheng},
  \citenamefont {Duan}, \citenamefont {Wang}, \citenamefont {Li}, \citenamefont
  {Deng},\ and\ \citenamefont {Wang}}]{PHE_tilt_2020}%
  \BibitemOpen
  \bibfield  {author} {\bibinfo {author} {\bibfnamefont {S.-H.}\ \bibnamefont
  {Zheng}}, \bibinfo {author} {\bibfnamefont {H.-J.}\ \bibnamefont {Duan}},
  \bibinfo {author} {\bibfnamefont {J.-K.}\ \bibnamefont {Wang}}, \bibinfo
  {author} {\bibfnamefont {J.-Y.}\ \bibnamefont {Li}}, \bibinfo {author}
  {\bibfnamefont {M.-X.}\ \bibnamefont {Deng}},\ and\ \bibinfo {author}
  {\bibfnamefont {R.-Q.}\ \bibnamefont {Wang}},\ }\bibfield  {title} {\bibinfo
  {title} {Origin of planar hall effect on the surface of topological
  insulators: Tilt of dirac cone by an in-plane magnetic field},\ }\href
  {https://doi.org/10.1103/PhysRevB.101.041408} {\bibfield  {journal} {\bibinfo
   {journal} {Phys. Rev. B}\ }\textbf {\bibinfo {volume} {101}},\ \bibinfo
  {pages} {041408} (\bibinfo {year} {2020})}\BibitemShut {NoStop}%
\bibitem [{\citenamefont {Zhang}\ \emph {et~al.}(2021)\citenamefont {Zhang},
  \citenamefont {Li}, \citenamefont {Pe\~na Benitez}, \citenamefont
  {Sur\'owka}, \citenamefont {Moessner}, \citenamefont {Molenkamp},\ and\
  \citenamefont {Trauzettel}}]{super_transport_2021}%
  \BibitemOpen
  \bibfield  {author} {\bibinfo {author} {\bibfnamefont {S.-B.}\ \bibnamefont
  {Zhang}}, \bibinfo {author} {\bibfnamefont {C.-A.}\ \bibnamefont {Li}},
  \bibinfo {author} {\bibfnamefont {F.}~\bibnamefont {Pe\~na Benitez}},
  \bibinfo {author} {\bibfnamefont {P.}~\bibnamefont {Sur\'owka}}, \bibinfo
  {author} {\bibfnamefont {R.}~\bibnamefont {Moessner}}, \bibinfo {author}
  {\bibfnamefont {L.~W.}\ \bibnamefont {Molenkamp}},\ and\ \bibinfo {author}
  {\bibfnamefont {B.}~\bibnamefont {Trauzettel}},\ }\bibfield  {title}
  {\bibinfo {title} {Super-resonant transport of topological surface states
  subjected to in-plane magnetic fields},\ }\href
  {https://doi.org/10.1103/PhysRevLett.127.076601} {\bibfield  {journal}
  {\bibinfo  {journal} {Phys. Rev. Lett.}\ }\textbf {\bibinfo {volume} {127}},\
  \bibinfo {pages} {076601} (\bibinfo {year} {2021})}\BibitemShut {NoStop}%
\bibitem [{SM()}]{SM}%
  \BibitemOpen
  \href@noop {} {}\bibinfo {note} {See Supplementary Material for the relevant
  discussions in detail, which includes
  Refs.~\cite{iteration_BTE_2001_prb,Ani_spin_2019,BC_role_2021,Liang_2009_warping,2010prb_fpstudies_cxLiu,NLPHE_2019_prl,Yu_nlpNernst_2021,2011prb_nonidealcone_YOichi,
  2009Sci_shen,2009Np_exp,tss_2010prl,Ando_2014prb,kimura_2018prb,Ashcroft_my,czeng_prr,2019NanoLett_DeltaT}}\BibitemShut
  {NoStop}%
\bibitem [{\citenamefont {Wang}\ and\ \citenamefont
  {Pang}(2010)}]{thermal_2010sscommu_WANG}%
  \BibitemOpen
  \bibfield  {author} {\bibinfo {author} {\bibfnamefont {C.}~\bibnamefont
  {Wang}}\ and\ \bibinfo {author} {\bibfnamefont {M.}~\bibnamefont {Pang}},\
  }\bibfield  {title} {\bibinfo {title} {Thermally induced spin polarization
  and thermal conductivities in a spin–orbit-coupled two-dimensional electron
  gas},\ }\href {https://doi.org/https://doi.org/10.1016/j.ssc.2010.06.013}
  {\bibfield  {journal} {\bibinfo  {journal} {Solid State Communications}\
  }\textbf {\bibinfo {volume} {150}},\ \bibinfo {pages} {1509} (\bibinfo {year}
  {2010})}\BibitemShut {NoStop}%
\bibitem [{\citenamefont {{Slachter}}\ \emph {et~al.}(2010)\citenamefont
  {{Slachter}}, \citenamefont {{Bakker}}, \citenamefont {{Adam}},\ and\
  \citenamefont {{van Wees}}}]{fm_spinInjection_2010_Nat.p_junction}%
  \BibitemOpen
  \bibfield  {author} {\bibinfo {author} {\bibfnamefont {A.}~\bibnamefont
  {{Slachter}}}, \bibinfo {author} {\bibfnamefont {F.~L.}\ \bibnamefont
  {{Bakker}}}, \bibinfo {author} {\bibfnamefont {J.~P.}\ \bibnamefont
  {{Adam}}},\ and\ \bibinfo {author} {\bibfnamefont {B.~J.}\ \bibnamefont {{van
  Wees}}},\ }\bibfield  {title} {\bibinfo {title} {{Thermally driven spin
  injection from a ferromagnet into a non-magnetic metal}},\ }\href
  {https://doi.org/10.1038/nphys1767} {\bibfield  {journal} {\bibinfo
  {journal} {Nature Physics}\ }\textbf {\bibinfo {volume} {6}},\ \bibinfo
  {pages} {879} (\bibinfo {year} {2010})}\BibitemShut {NoStop}%
\bibitem [{\citenamefont {{Le Breton}}\ \emph {et~al.}(2011)\citenamefont {{Le
  Breton}}, \citenamefont {{Sharma}}, \citenamefont {{Saito}}, \citenamefont
  {{Yuasa}},\ and\ \citenamefont {{Jansen}}}]{fm_spinC_2011_Nat_junction}%
  \BibitemOpen
  \bibfield  {author} {\bibinfo {author} {\bibfnamefont {J.-C.}\ \bibnamefont
  {{Le Breton}}}, \bibinfo {author} {\bibfnamefont {S.}~\bibnamefont
  {{Sharma}}}, \bibinfo {author} {\bibfnamefont {H.}~\bibnamefont {{Saito}}},
  \bibinfo {author} {\bibfnamefont {S.}~\bibnamefont {{Yuasa}}},\ and\ \bibinfo
  {author} {\bibfnamefont {R.}~\bibnamefont {{Jansen}}},\ }\bibfield  {title}
  {\bibinfo {title} {{Thermal spin current from a ferromagnet to silicon by
  Seebeck spin tunnelling}},\ }\href {https://doi.org/10.1038/nature10224}
  {\bibfield  {journal} {\bibinfo  {journal} {Nature}\ }\textbf {\bibinfo
  {volume} {475}},\ \bibinfo {pages} {82} (\bibinfo {year} {2011})}\BibitemShut
  {NoStop}%
\bibitem [{\citenamefont {Dyrda\l{}}\ \emph {et~al.}(2013)\citenamefont
  {Dyrda\l{}}, \citenamefont {Inglot}, \citenamefont {Dugaev},\ and\
  \citenamefont {Barna\ifmmode~\acute{s}\else
  \'{s}\fi{}}}]{thermal_2013prb_dyrdal}%
  \BibitemOpen
  \bibfield  {author} {\bibinfo {author} {\bibfnamefont {A.}~\bibnamefont
  {Dyrda\l{}}}, \bibinfo {author} {\bibfnamefont {M.}~\bibnamefont {Inglot}},
  \bibinfo {author} {\bibfnamefont {V.~K.}\ \bibnamefont {Dugaev}},\ and\
  \bibinfo {author} {\bibfnamefont {J.}~\bibnamefont
  {Barna\ifmmode~\acute{s}\else \'{s}\fi{}}},\ }\bibfield  {title} {\bibinfo
  {title} {Thermally induced spin polarization of a two-dimensional electron
  gas},\ }\href {https://doi.org/10.1103/PhysRevB.87.245309} {\bibfield
  {journal} {\bibinfo  {journal} {Phys. Rev. B}\ }\textbf {\bibinfo {volume}
  {87}},\ \bibinfo {pages} {245309} (\bibinfo {year} {2013})}\BibitemShut
  {NoStop}%
\bibitem [{\citenamefont {Xiao}\ \emph {et~al.}(2016)\citenamefont {Xiao},
  \citenamefont {Li},\ and\ \citenamefont
  {Ma}}]{thermal_2016frontiers_zhongshui}%
  \BibitemOpen
  \bibfield  {author} {\bibinfo {author} {\bibfnamefont {C.}~\bibnamefont
  {Xiao}}, \bibinfo {author} {\bibfnamefont {D.}~\bibnamefont {Li}},\ and\
  \bibinfo {author} {\bibfnamefont {Z.}~\bibnamefont {Ma}},\ }\bibfield
  {title} {\bibinfo {title} {Thermoelectric response of spin polarization in
  rashba spintronic systems},\ }\href@noop {} {\bibfield  {journal} {\bibinfo
  {journal} {Frontiers of Physics}\ }\textbf {\bibinfo {volume} {11}},\
  \bibinfo {pages} {1} (\bibinfo {year} {2016})}\BibitemShut {NoStop}%
\bibitem [{\citenamefont {Dyrda\l{}}\ \emph {et~al.}(2018)\citenamefont
  {Dyrda\l{}}, \citenamefont {Barna\ifmmode~\acute{s}\else \'{s}\fi{}},
  \citenamefont {Dugaev},\ and\ \citenamefont
  {Berakdar}}]{thermal_2018prb_dyrdal}%
  \BibitemOpen
  \bibfield  {author} {\bibinfo {author} {\bibfnamefont {A.}~\bibnamefont
  {Dyrda\l{}}}, \bibinfo {author} {\bibfnamefont {J.}~\bibnamefont
  {Barna\ifmmode~\acute{s}\else \'{s}\fi{}}}, \bibinfo {author} {\bibfnamefont
  {V.~K.}\ \bibnamefont {Dugaev}},\ and\ \bibinfo {author} {\bibfnamefont
  {J.}~\bibnamefont {Berakdar}},\ }\bibfield  {title} {\bibinfo {title}
  {Thermally induced spin polarization in a magnetized two-dimensional electron
  gas with rashba spin-orbit interaction},\ }\href
  {https://doi.org/10.1103/PhysRevB.98.075307} {\bibfield  {journal} {\bibinfo
  {journal} {Phys. Rev. B}\ }\textbf {\bibinfo {volume} {98}},\ \bibinfo
  {pages} {075307} (\bibinfo {year} {2018})}\BibitemShut {NoStop}%
\bibitem [{\citenamefont {Ning}\ \emph {et~al.}(2021)\citenamefont {Ning},
  \citenamefont {Peng}, \citenamefont {Wang}, \citenamefont {Chen},
  \citenamefont {Yang}, \citenamefont {Chen}, \citenamefont {Zhao},
  \citenamefont {Sun}, \citenamefont {Kanagaraj}, \citenamefont {Yang},
  \citenamefont {Gao}, \citenamefont {Zhang}, \citenamefont {Zhao},
  \citenamefont {Pan}, \citenamefont {Ruan}, \citenamefont {Li}, \citenamefont
  {Liu}, \citenamefont {He}, \citenamefont {Chen},\ and\ \citenamefont
  {Xu}}]{spc_2021_prb}%
  \BibitemOpen
  \bibfield  {author} {\bibinfo {author} {\bibfnamefont {J.}~\bibnamefont
  {Ning}}, \bibinfo {author} {\bibfnamefont {W.}~\bibnamefont {Peng}}, \bibinfo
  {author} {\bibfnamefont {W.}~\bibnamefont {Wang}}, \bibinfo {author}
  {\bibfnamefont {Z.}~\bibnamefont {Chen}}, \bibinfo {author} {\bibfnamefont
  {P.}~\bibnamefont {Yang}}, \bibinfo {author} {\bibfnamefont {Y.}~\bibnamefont
  {Chen}}, \bibinfo {author} {\bibfnamefont {Y.}~\bibnamefont {Zhao}}, \bibinfo
  {author} {\bibfnamefont {Y.}~\bibnamefont {Sun}}, \bibinfo {author}
  {\bibfnamefont {M.}~\bibnamefont {Kanagaraj}}, \bibinfo {author}
  {\bibfnamefont {L.}~\bibnamefont {Yang}}, \bibinfo {author} {\bibfnamefont
  {Q.}~\bibnamefont {Gao}}, \bibinfo {author} {\bibfnamefont {J.}~\bibnamefont
  {Zhang}}, \bibinfo {author} {\bibfnamefont {D.}~\bibnamefont {Zhao}},
  \bibinfo {author} {\bibfnamefont {D.}~\bibnamefont {Pan}}, \bibinfo {author}
  {\bibfnamefont {X.}~\bibnamefont {Ruan}}, \bibinfo {author} {\bibfnamefont
  {Y.}~\bibnamefont {Li}}, \bibinfo {author} {\bibfnamefont {W.}~\bibnamefont
  {Liu}}, \bibinfo {author} {\bibfnamefont {L.}~\bibnamefont {He}}, \bibinfo
  {author} {\bibfnamefont {Z.-G.}\ \bibnamefont {Chen}},\ and\ \bibinfo
  {author} {\bibfnamefont {Y.}~\bibnamefont {Xu}},\ }\bibfield  {title}
  {\bibinfo {title} {Thermal induced spin-polarized current protected by
  spin-momentum locking in ${\mathrm{zrte}}_{5}$ nanowires},\ }\href
  {https://doi.org/10.1103/PhysRevB.104.035429} {\bibfield  {journal} {\bibinfo
   {journal} {Phys. Rev. B}\ }\textbf {\bibinfo {volume} {104}},\ \bibinfo
  {pages} {035429} (\bibinfo {year} {2021})}\BibitemShut {NoStop}%
\bibitem [{\citenamefont {Wang}\ \emph {et~al.}(2001)\citenamefont {Wang},
  \citenamefont {Xu}, \citenamefont {Kakeshita}, \citenamefont {Uchida},
  \citenamefont {Ono}, \citenamefont {Ando},\ and\ \citenamefont
  {Ong}}]{iteration_BTE_2001_prb}%
  \BibitemOpen
  \bibfield  {author} {\bibinfo {author} {\bibfnamefont {Y.}~\bibnamefont
  {Wang}}, \bibinfo {author} {\bibfnamefont {Z.~A.}\ \bibnamefont {Xu}},
  \bibinfo {author} {\bibfnamefont {T.}~\bibnamefont {Kakeshita}}, \bibinfo
  {author} {\bibfnamefont {S.}~\bibnamefont {Uchida}}, \bibinfo {author}
  {\bibfnamefont {S.}~\bibnamefont {Ono}}, \bibinfo {author} {\bibfnamefont
  {Y.}~\bibnamefont {Ando}},\ and\ \bibinfo {author} {\bibfnamefont {N.~P.}\
  \bibnamefont {Ong}},\ }\bibfield  {title} {\bibinfo {title} {Onset of the
  vortexlike nernst signal above ${T}_{c}$ in
  ${\mathrm{la}}_{2\ensuremath{-}x}{\mathrm{sr}}_{x}{\mathrm{cuo}}_{4}$ and
  ${\mathrm{bi}}_{2}{\mathrm{sr}}_{2\ensuremath{-}y}{\mathrm{la}}_{y}{\mathrm{cuo}}_{6}$},\
  }\href {https://doi.org/10.1103/PhysRevB.64.224519} {\bibfield  {journal}
  {\bibinfo  {journal} {Phys. Rev. B}\ }\textbf {\bibinfo {volume} {64}},\
  \bibinfo {pages} {224519} (\bibinfo {year} {2001})}\BibitemShut {NoStop}%
\bibitem [{\citenamefont {Xiao}\ \emph {et~al.}(2010)\citenamefont {Xiao},
  \citenamefont {Chang},\ and\ \citenamefont {Niu}}]{Beffect_2010_niu}%
  \BibitemOpen
  \bibfield  {author} {\bibinfo {author} {\bibfnamefont {D.}~\bibnamefont
  {Xiao}}, \bibinfo {author} {\bibfnamefont {M.-C.}\ \bibnamefont {Chang}},\
  and\ \bibinfo {author} {\bibfnamefont {Q.}~\bibnamefont {Niu}},\ }\bibfield
  {title} {\bibinfo {title} {Berry phase effects on electronic properties},\
  }\href {https://doi.org/10.1103/RevModPhys.82.1959} {\bibfield  {journal}
  {\bibinfo  {journal} {Rev. Mod. Phys.}\ }\textbf {\bibinfo {volume} {82}},\
  \bibinfo {pages} {1959} (\bibinfo {year} {2010})}\BibitemShut {NoStop}%
\bibitem [{\citenamefont {Qin}\ \emph {et~al.}(2011)\citenamefont {Qin},
  \citenamefont {Niu},\ and\ \citenamefont {Shi}}]{heatcurrent_2011_niu}%
  \BibitemOpen
  \bibfield  {author} {\bibinfo {author} {\bibfnamefont {T.}~\bibnamefont
  {Qin}}, \bibinfo {author} {\bibfnamefont {Q.}~\bibnamefont {Niu}},\ and\
  \bibinfo {author} {\bibfnamefont {J.}~\bibnamefont {Shi}},\ }\bibfield
  {title} {\bibinfo {title} {Energy magnetization and the thermal hall
  effect},\ }\href {https://doi.org/10.1103/PhysRevLett.107.236601} {\bibfield
  {journal} {\bibinfo  {journal} {Phys. Rev. Lett.}\ }\textbf {\bibinfo
  {volume} {107}},\ \bibinfo {pages} {236601} (\bibinfo {year}
  {2011})}\BibitemShut {NoStop}%
\bibitem [{\citenamefont {Yang}\ \emph {et~al.}(2018)\citenamefont {Yang},
  \citenamefont {Wang},\ and\ \citenamefont {Liu}}]{interplay_2018}%
  \BibitemOpen
  \bibfield  {author} {\bibinfo {author} {\bibfnamefont {Z.-K.}\ \bibnamefont
  {Yang}}, \bibinfo {author} {\bibfnamefont {J.-R.}\ \bibnamefont {Wang}},\
  and\ \bibinfo {author} {\bibfnamefont {G.-Z.}\ \bibnamefont {Liu}},\
  }\bibfield  {title} {\bibinfo {title} {Effects of dirac cone tilt in a
  two-dimensional dirac semimetal},\ }\href
  {https://doi.org/10.1103/PhysRevB.98.195123} {\bibfield  {journal} {\bibinfo
  {journal} {Phys. Rev. B}\ }\textbf {\bibinfo {volume} {98}},\ \bibinfo
  {pages} {195123} (\bibinfo {year} {2018})}\BibitemShut {NoStop}%
\bibitem [{\citenamefont {Zhao}\ and\ \citenamefont
  {Wang}(2019)}]{interplay_2019}%
  \BibitemOpen
  \bibfield  {author} {\bibinfo {author} {\bibfnamefont {P.-L.}\ \bibnamefont
  {Zhao}}\ and\ \bibinfo {author} {\bibfnamefont {A.-M.}\ \bibnamefont
  {Wang}},\ }\bibfield  {title} {\bibinfo {title} {Interplay between tilt,
  disorder, and coulomb interaction in type-i dirac fermions},\ }\href
  {https://doi.org/10.1103/PhysRevB.100.125138} {\bibfield  {journal} {\bibinfo
   {journal} {Phys. Rev. B}\ }\textbf {\bibinfo {volume} {100}},\ \bibinfo
  {pages} {125138} (\bibinfo {year} {2019})}\BibitemShut {NoStop}%
\bibitem [{\citenamefont {Choi}\ \emph {et~al.}(2010)\citenamefont {Choi},
  \citenamefont {Jhi},\ and\ \citenamefont {Son}}]{tilted_graphene_2010}%
  \BibitemOpen
  \bibfield  {author} {\bibinfo {author} {\bibfnamefont {S.-M.}\ \bibnamefont
  {Choi}}, \bibinfo {author} {\bibfnamefont {S.-H.}\ \bibnamefont {Jhi}},\ and\
  \bibinfo {author} {\bibfnamefont {Y.-W.}\ \bibnamefont {Son}},\ }\bibfield
  {title} {\bibinfo {title} {Effects of strain on electronic properties of
  graphene},\ }\href {https://doi.org/10.1103/PhysRevB.81.081407} {\bibfield
  {journal} {\bibinfo  {journal} {Phys. Rev. B}\ }\textbf {\bibinfo {volume}
  {81}},\ \bibinfo {pages} {081407} (\bibinfo {year} {2010})}\BibitemShut
  {NoStop}%
\bibitem [{\citenamefont {Goerbig}\ \emph {et~al.}(2008)\citenamefont
  {Goerbig}, \citenamefont {Fuchs}, \citenamefont {Montambaux},\ and\
  \citenamefont {Pi\'echon}}]{tilted_DSM_graphene_2008}%
  \BibitemOpen
  \bibfield  {author} {\bibinfo {author} {\bibfnamefont {M.~O.}\ \bibnamefont
  {Goerbig}}, \bibinfo {author} {\bibfnamefont {J.-N.}\ \bibnamefont {Fuchs}},
  \bibinfo {author} {\bibfnamefont {G.}~\bibnamefont {Montambaux}},\ and\
  \bibinfo {author} {\bibfnamefont {F.}~\bibnamefont {Pi\'echon}},\ }\bibfield
  {title} {\bibinfo {title} {Tilted anisotropic dirac cones in quinoid-type
  graphene and
  $\ensuremath{\alpha}\text{\ensuremath{-}}{(\text{BEDT-TTF})}_{2}{\text{i}}_{3}$},\
  }\href {https://doi.org/10.1103/PhysRevB.78.045415} {\bibfield  {journal}
  {\bibinfo  {journal} {Phys. Rev. B}\ }\textbf {\bibinfo {volume} {78}},\
  \bibinfo {pages} {045415} (\bibinfo {year} {2008})}\BibitemShut {NoStop}%
\bibitem [{\citenamefont {Tanaka}\ \emph {et~al.}(2012)\citenamefont {Tanaka},
  \citenamefont {Ren}, \citenamefont {Sato}, \citenamefont {Nakayama},
  \citenamefont {Souma}, \citenamefont {Takahashi}, \citenamefont {Segawa},\
  and\ \citenamefont {Ando}}]{TCI_2012}%
  \BibitemOpen
  \bibfield  {author} {\bibinfo {author} {\bibfnamefont {Y.}~\bibnamefont
  {Tanaka}}, \bibinfo {author} {\bibfnamefont {Z.}~\bibnamefont {Ren}},
  \bibinfo {author} {\bibfnamefont {T.}~\bibnamefont {Sato}}, \bibinfo {author}
  {\bibfnamefont {K.}~\bibnamefont {Nakayama}}, \bibinfo {author}
  {\bibfnamefont {S.}~\bibnamefont {Souma}}, \bibinfo {author} {\bibfnamefont
  {T.}~\bibnamefont {Takahashi}}, \bibinfo {author} {\bibfnamefont
  {K.}~\bibnamefont {Segawa}},\ and\ \bibinfo {author} {\bibfnamefont
  {Y.}~\bibnamefont {Ando}},\ }\bibfield  {title} {\bibinfo {title}
  {Experimental realization of a topological crystalline insulator in snte},\
  }\href {https://doi.org/10.1038/nphys2442} {\bibfield  {journal} {\bibinfo
  {journal} {Nature Physics}\ }\textbf {\bibinfo {volume} {8}},\ \bibinfo
  {pages} {800–803} (\bibinfo {year} {2012})}\BibitemShut {NoStop}%
\bibitem [{\citenamefont {Sodemann}\ and\ \citenamefont
  {Fu}(2015)}]{TCI_Inti_2015}%
  \BibitemOpen
  \bibfield  {author} {\bibinfo {author} {\bibfnamefont {I.}~\bibnamefont
  {Sodemann}}\ and\ \bibinfo {author} {\bibfnamefont {L.}~\bibnamefont {Fu}},\
  }\bibfield  {title} {\bibinfo {title} {Quantum nonlinear hall effect induced
  by berry curvature dipole in time-reversal invariant materials},\ }\href
  {https://doi.org/10.1103/PhysRevLett.115.216806} {\bibfield  {journal}
  {\bibinfo  {journal} {Phys. Rev. Lett.}\ }\textbf {\bibinfo {volume} {115}},\
  \bibinfo {pages} {216806} (\bibinfo {year} {2015})}\BibitemShut {NoStop}%
\bibitem [{\citenamefont {Chiu}\ \emph {et~al.}(2017)\citenamefont {Chiu},
  \citenamefont {Chan}, \citenamefont {Li}, \citenamefont {Nohara},\ and\
  \citenamefont {Schnyder}}]{TCI_2017}%
  \BibitemOpen
  \bibfield  {author} {\bibinfo {author} {\bibfnamefont {C.-K.}\ \bibnamefont
  {Chiu}}, \bibinfo {author} {\bibfnamefont {Y.-H.}\ \bibnamefont {Chan}},
  \bibinfo {author} {\bibfnamefont {X.}~\bibnamefont {Li}}, \bibinfo {author}
  {\bibfnamefont {Y.}~\bibnamefont {Nohara}},\ and\ \bibinfo {author}
  {\bibfnamefont {A.~P.}\ \bibnamefont {Schnyder}},\ }\bibfield  {title}
  {\bibinfo {title} {Type-ii dirac surface states in topological crystalline
  insulators},\ }\href {https://doi.org/10.1103/PhysRevB.95.035151} {\bibfield
  {journal} {\bibinfo  {journal} {Phys. Rev. B}\ }\textbf {\bibinfo {volume}
  {95}},\ \bibinfo {pages} {035151} (\bibinfo {year} {2017})}\BibitemShut
  {NoStop}%
\bibitem [{\citenamefont {Liu}\ \emph {et~al.}(2010)\citenamefont {Liu},
  \citenamefont {Qi}, \citenamefont {Zhang}, \citenamefont {Dai}, \citenamefont
  {Fang},\ and\ \citenamefont {Zhang}}]{2010prb_fpstudies_cxLiu}%
  \BibitemOpen
  \bibfield  {author} {\bibinfo {author} {\bibfnamefont {C.-X.}\ \bibnamefont
  {Liu}}, \bibinfo {author} {\bibfnamefont {X.-L.}\ \bibnamefont {Qi}},
  \bibinfo {author} {\bibfnamefont {H.}~\bibnamefont {Zhang}}, \bibinfo
  {author} {\bibfnamefont {X.}~\bibnamefont {Dai}}, \bibinfo {author}
  {\bibfnamefont {Z.}~\bibnamefont {Fang}},\ and\ \bibinfo {author}
  {\bibfnamefont {S.-C.}\ \bibnamefont {Zhang}},\ }\bibfield  {title} {\bibinfo
  {title} {Model hamiltonian for topological insulators},\ }\href
  {https://doi.org/10.1103/PhysRevB.82.045122} {\bibfield  {journal} {\bibinfo
  {journal} {Phys. Rev. B}\ }\textbf {\bibinfo {volume} {82}},\ \bibinfo
  {pages} {045122} (\bibinfo {year} {2010})}\BibitemShut {NoStop}%
\bibitem [{\citenamefont {Taskin}\ and\ \citenamefont
  {Ando}(2011)}]{2011prb_nonidealcone_YOichi}%
  \BibitemOpen
  \bibfield  {author} {\bibinfo {author} {\bibfnamefont {A.~A.}\ \bibnamefont
  {Taskin}}\ and\ \bibinfo {author} {\bibfnamefont {Y.}~\bibnamefont {Ando}},\
  }\bibfield  {title} {\bibinfo {title} {Berry phase of nonideal dirac fermions
  in topological insulators},\ }\href
  {https://doi.org/10.1103/PhysRevB.84.035301} {\bibfield  {journal} {\bibinfo
  {journal} {Phys. Rev. B}\ }\textbf {\bibinfo {volume} {84}},\ \bibinfo
  {pages} {035301} (\bibinfo {year} {2011})}\BibitemShut {NoStop}%
\bibitem [{\citenamefont {{Chen}}\ \emph {et~al.}(2009)\citenamefont {{Chen}},
  \citenamefont {{Analytis}}, \citenamefont {{Chu}}, \citenamefont {{Liu}},
  \citenamefont {{Mo}}, \citenamefont {{Qi}}, \citenamefont {{Zhang}},
  \citenamefont {{Lu}}, \citenamefont {{Dai}}, \citenamefont {{Fang}},
  \citenamefont {{Zhang}}, \citenamefont {{Fisher}}, \citenamefont
  {{Hussain}},\ and\ \citenamefont {{Shen}}}]{2009Sci_shen}%
  \BibitemOpen
  \bibfield  {author} {\bibinfo {author} {\bibfnamefont {Y.~L.}\ \bibnamefont
  {{Chen}}}, \bibinfo {author} {\bibfnamefont {J.~G.}\ \bibnamefont
  {{Analytis}}}, \bibinfo {author} {\bibfnamefont {J.~H.}\ \bibnamefont
  {{Chu}}}, \bibinfo {author} {\bibfnamefont {Z.~K.}\ \bibnamefont {{Liu}}},
  \bibinfo {author} {\bibfnamefont {S.~K.}\ \bibnamefont {{Mo}}}, \bibinfo
  {author} {\bibfnamefont {X.~L.}\ \bibnamefont {{Qi}}}, \bibinfo {author}
  {\bibfnamefont {H.~J.}\ \bibnamefont {{Zhang}}}, \bibinfo {author}
  {\bibfnamefont {D.~H.}\ \bibnamefont {{Lu}}}, \bibinfo {author}
  {\bibfnamefont {X.}~\bibnamefont {{Dai}}}, \bibinfo {author} {\bibfnamefont
  {Z.}~\bibnamefont {{Fang}}}, \bibinfo {author} {\bibfnamefont {S.~C.}\
  \bibnamefont {{Zhang}}}, \bibinfo {author} {\bibfnamefont {I.~R.}\
  \bibnamefont {{Fisher}}}, \bibinfo {author} {\bibfnamefont {Z.}~\bibnamefont
  {{Hussain}}},\ and\ \bibinfo {author} {\bibfnamefont {Z.~X.}\ \bibnamefont
  {{Shen}}},\ }\bibfield  {title} {\bibinfo {title} {{Experimental Realization
  of a Three-Dimensional Topological Insulator, Bi$_{2}$Te$_{3}$}},\ }\href
  {https://doi.org/10.1126/science.1173034} {\bibfield  {journal} {\bibinfo
  {journal} {Science}\ }\textbf {\bibinfo {volume} {325}},\ \bibinfo {pages}
  {178} (\bibinfo {year} {2009})}\BibitemShut {NoStop}%
\bibitem [{\citenamefont {{Xia}}\ \emph {et~al.}(2009)\citenamefont {{Xia}},
  \citenamefont {{Qian}}, \citenamefont {{Hsieh}}, \citenamefont {{Wray}},
  \citenamefont {{Pal}}, \citenamefont {{Lin}}, \citenamefont {{Bansil}},
  \citenamefont {{Grauer}}, \citenamefont {{Hor}}, \citenamefont {{Cava}},\
  and\ \citenamefont {{Hasan}}}]{2009Np_exp}%
  \BibitemOpen
  \bibfield  {author} {\bibinfo {author} {\bibfnamefont {Y.}~\bibnamefont
  {{Xia}}}, \bibinfo {author} {\bibfnamefont {D.}~\bibnamefont {{Qian}}},
  \bibinfo {author} {\bibfnamefont {D.}~\bibnamefont {{Hsieh}}}, \bibinfo
  {author} {\bibfnamefont {L.}~\bibnamefont {{Wray}}}, \bibinfo {author}
  {\bibfnamefont {A.}~\bibnamefont {{Pal}}}, \bibinfo {author} {\bibfnamefont
  {H.}~\bibnamefont {{Lin}}}, \bibinfo {author} {\bibfnamefont
  {A.}~\bibnamefont {{Bansil}}}, \bibinfo {author} {\bibfnamefont
  {D.}~\bibnamefont {{Grauer}}}, \bibinfo {author} {\bibfnamefont {Y.~S.}\
  \bibnamefont {{Hor}}}, \bibinfo {author} {\bibfnamefont {R.~J.}\ \bibnamefont
  {{Cava}}},\ and\ \bibinfo {author} {\bibfnamefont {M.~Z.}\ \bibnamefont
  {{Hasan}}},\ }\bibfield  {title} {\bibinfo {title} {{Observation of a
  large-gap topological-insulator class with a single Dirac cone on the
  surface}},\ }\href {https://doi.org/10.1038/nphys1274} {\bibfield  {journal}
  {\bibinfo  {journal} {Nature Physics}\ }\textbf {\bibinfo {volume} {5}},\
  \bibinfo {pages} {398} (\bibinfo {year} {2009})}\BibitemShut {NoStop}%
\bibitem [{\citenamefont {Alpichshev}\ \emph {et~al.}(2010)\citenamefont
  {Alpichshev}, \citenamefont {Analytis}, \citenamefont {Chu}, \citenamefont
  {Fisher}, \citenamefont {Chen}, \citenamefont {Shen}, \citenamefont {Fang},\
  and\ \citenamefont {Kapitulnik}}]{tss_2010prl}%
  \BibitemOpen
  \bibfield  {author} {\bibinfo {author} {\bibfnamefont {Z.}~\bibnamefont
  {Alpichshev}}, \bibinfo {author} {\bibfnamefont {J.~G.}\ \bibnamefont
  {Analytis}}, \bibinfo {author} {\bibfnamefont {J.-H.}\ \bibnamefont {Chu}},
  \bibinfo {author} {\bibfnamefont {I.~R.}\ \bibnamefont {Fisher}}, \bibinfo
  {author} {\bibfnamefont {Y.~L.}\ \bibnamefont {Chen}}, \bibinfo {author}
  {\bibfnamefont {Z.~X.}\ \bibnamefont {Shen}}, \bibinfo {author}
  {\bibfnamefont {A.}~\bibnamefont {Fang}},\ and\ \bibinfo {author}
  {\bibfnamefont {A.}~\bibnamefont {Kapitulnik}},\ }\bibfield  {title}
  {\bibinfo {title} {Stm imaging of electronic waves on the surface of
  ${\mathrm{bi}}_{2}{\mathrm{te}}_{3}$: Topologically protected surface states
  and hexagonal warping effects},\ }\href
  {https://doi.org/10.1103/PhysRevLett.104.016401} {\bibfield  {journal}
  {\bibinfo  {journal} {Phys. Rev. Lett.}\ }\textbf {\bibinfo {volume} {104}},\
  \bibinfo {pages} {016401} (\bibinfo {year} {2010})}\BibitemShut {NoStop}%
\bibitem [{\citenamefont {Nomura}\ \emph {et~al.}(2014)\citenamefont {Nomura},
  \citenamefont {Souma}, \citenamefont {Takayama}, \citenamefont {Sato},
  \citenamefont {Takahashi}, \citenamefont {Eto}, \citenamefont {Segawa},\ and\
  \citenamefont {Ando}}]{Ando_2014prb}%
  \BibitemOpen
  \bibfield  {author} {\bibinfo {author} {\bibfnamefont {M.}~\bibnamefont
  {Nomura}}, \bibinfo {author} {\bibfnamefont {S.}~\bibnamefont {Souma}},
  \bibinfo {author} {\bibfnamefont {A.}~\bibnamefont {Takayama}}, \bibinfo
  {author} {\bibfnamefont {T.}~\bibnamefont {Sato}}, \bibinfo {author}
  {\bibfnamefont {T.}~\bibnamefont {Takahashi}}, \bibinfo {author}
  {\bibfnamefont {K.}~\bibnamefont {Eto}}, \bibinfo {author} {\bibfnamefont
  {K.}~\bibnamefont {Segawa}},\ and\ \bibinfo {author} {\bibfnamefont
  {Y.}~\bibnamefont {Ando}},\ }\bibfield  {title} {\bibinfo {title}
  {Relationship between fermi surface warping and out-of-plane spin
  polarization in topological insulators: A view from spin- and angle-resolved
  photoemission},\ }\href {https://doi.org/10.1103/PhysRevB.89.045134}
  {\bibfield  {journal} {\bibinfo  {journal} {Phys. Rev. B}\ }\textbf {\bibinfo
  {volume} {89}},\ \bibinfo {pages} {045134} (\bibinfo {year}
  {2014})}\BibitemShut {NoStop}%
\bibitem [{\citenamefont {Annese}\ \emph {et~al.}(2018)\citenamefont {Annese},
  \citenamefont {Okuda}, \citenamefont {Schwier}, \citenamefont {Iwasawa},
  \citenamefont {Shimada}, \citenamefont {Natamane}, \citenamefont {Taniguchi},
  \citenamefont {Rusinov}, \citenamefont {Eremeev}, \citenamefont {Kokh},
  \citenamefont {Golyashov}, \citenamefont {Tereshchenko}, \citenamefont
  {Chulkov},\ and\ \citenamefont {Kimura}}]{kimura_2018prb}%
  \BibitemOpen
  \bibfield  {author} {\bibinfo {author} {\bibfnamefont {E.}~\bibnamefont
  {Annese}}, \bibinfo {author} {\bibfnamefont {T.}~\bibnamefont {Okuda}},
  \bibinfo {author} {\bibfnamefont {E.~F.}\ \bibnamefont {Schwier}}, \bibinfo
  {author} {\bibfnamefont {H.}~\bibnamefont {Iwasawa}}, \bibinfo {author}
  {\bibfnamefont {K.}~\bibnamefont {Shimada}}, \bibinfo {author} {\bibfnamefont
  {M.}~\bibnamefont {Natamane}}, \bibinfo {author} {\bibfnamefont
  {M.}~\bibnamefont {Taniguchi}}, \bibinfo {author} {\bibfnamefont {I.~P.}\
  \bibnamefont {Rusinov}}, \bibinfo {author} {\bibfnamefont {S.~V.}\
  \bibnamefont {Eremeev}}, \bibinfo {author} {\bibfnamefont {K.~A.}\
  \bibnamefont {Kokh}}, \bibinfo {author} {\bibfnamefont {V.~A.}\ \bibnamefont
  {Golyashov}}, \bibinfo {author} {\bibfnamefont {O.~E.}\ \bibnamefont
  {Tereshchenko}}, \bibinfo {author} {\bibfnamefont {E.~V.}\ \bibnamefont
  {Chulkov}},\ and\ \bibinfo {author} {\bibfnamefont {A.}~\bibnamefont
  {Kimura}},\ }\bibfield  {title} {\bibinfo {title} {Electronic and spin
  structure of the wide-band-gap topological insulator: Nearly stoichiometric
  ${\mathbf{bi}}_{2}{\mathbf{te}}_{2}\mathbf{S}$},\ }\href
  {https://doi.org/10.1103/PhysRevB.97.205113} {\bibfield  {journal} {\bibinfo
  {journal} {Phys. Rev. B}\ }\textbf {\bibinfo {volume} {97}},\ \bibinfo
  {pages} {205113} (\bibinfo {year} {2018})}\BibitemShut {NoStop}%
\bibitem [{\citenamefont {{Xu}}\ \emph {et~al.}(2019)\citenamefont {{Xu}},
  \citenamefont {{Phelan}},\ and\ \citenamefont
  {{Chien}}}]{2019NanoLett_DeltaT}%
  \BibitemOpen
  \bibfield  {author} {\bibinfo {author} {\bibfnamefont {J.}~\bibnamefont
  {{Xu}}}, \bibinfo {author} {\bibfnamefont {W.~A.}\ \bibnamefont {{Phelan}}},\
  and\ \bibinfo {author} {\bibfnamefont {C.-L.}\ \bibnamefont {{Chien}}},\
  }\bibfield  {title} {\bibinfo {title} {{Large Anomalous Nernst Effect in a
  van der Waals Ferromagnet Fe3GeTe2}},\ }\href
  {https://doi.org/10.1021/acs.nanolett.9b03739} {\bibfield  {journal}
  {\bibinfo  {journal} {Nano Letters}\ }\textbf {\bibinfo {volume} {19}},\
  \bibinfo {pages} {8250} (\bibinfo {year} {2019})}\BibitemShut {NoStop}%
\bibitem [{\citenamefont {Steinberg}\ \emph {et~al.}(2010)\citenamefont
  {Steinberg}, \citenamefont {Gardner}, \citenamefont {Lee},\ and\
  \citenamefont {Jarillo-Herrero}}]{2010Nanolett_Pablo}%
  \BibitemOpen
  \bibfield  {author} {\bibinfo {author} {\bibfnamefont {H.}~\bibnamefont
  {Steinberg}}, \bibinfo {author} {\bibfnamefont {D.~R.}\ \bibnamefont
  {Gardner}}, \bibinfo {author} {\bibfnamefont {Y.~S.}\ \bibnamefont {Lee}},\
  and\ \bibinfo {author} {\bibfnamefont {P.}~\bibnamefont {Jarillo-Herrero}},\
  }\bibfield  {title} {\bibinfo {title} {Surface state transport and ambipolar
  electric field effect in bi$_2$se$_3$ nanodevices},\ }\href
  {https://doi.org/10.1021/nl1032183} {\bibfield  {journal} {\bibinfo
  {journal} {Nano Letters}\ }\textbf {\bibinfo {volume} {10}},\ \bibinfo
  {pages} {5032} (\bibinfo {year} {2010})}\BibitemShut {NoStop}%
\bibitem [{\citenamefont {{Hong}}\ \emph {et~al.}(2012)\citenamefont {{Hong}},
  \citenamefont {{Cha}}, \citenamefont {{Kong}},\ and\ \citenamefont
  {{Cui}}}]{2012NC_cui}%
  \BibitemOpen
  \bibfield  {author} {\bibinfo {author} {\bibfnamefont {S.~S.}\ \bibnamefont
  {{Hong}}}, \bibinfo {author} {\bibfnamefont {J.~J.}\ \bibnamefont {{Cha}}},
  \bibinfo {author} {\bibfnamefont {D.}~\bibnamefont {{Kong}}},\ and\ \bibinfo
  {author} {\bibfnamefont {Y.}~\bibnamefont {{Cui}}},\ }\bibfield  {title}
  {\bibinfo {title} {{Ultra-low carrier concentration and surface-dominant
  transport in antimony-doped Bi$_{2}$Se$_{3}$ topological insulator
  nanoribbons}},\ }\href {https://doi.org/10.1038/ncomms1771} {\bibfield
  {journal} {\bibinfo  {journal} {Nature Communications}\ }\textbf {\bibinfo
  {volume} {3}},\ \bibinfo {eid} {757} (\bibinfo {year} {2012})}\BibitemShut
  {NoStop}%
\bibitem [{\citenamefont {{Yoshimi}}\ \emph {et~al.}(2015)\citenamefont
  {{Yoshimi}}, \citenamefont {{Tsukazaki}}, \citenamefont {{Kozuka}},
  \citenamefont {{Falson}}, \citenamefont {{Takahashi}}, \citenamefont
  {{Checkelsky}}, \citenamefont {{Nagaosa}}, \citenamefont {{Kawasaki}},\ and\
  \citenamefont {{Tokura}}}]{2015NC_Tokura}%
  \BibitemOpen
  \bibfield  {author} {\bibinfo {author} {\bibfnamefont {R.}~\bibnamefont
  {{Yoshimi}}}, \bibinfo {author} {\bibfnamefont {A.}~\bibnamefont
  {{Tsukazaki}}}, \bibinfo {author} {\bibfnamefont {Y.}~\bibnamefont
  {{Kozuka}}}, \bibinfo {author} {\bibfnamefont {J.}~\bibnamefont {{Falson}}},
  \bibinfo {author} {\bibfnamefont {K.~S.}\ \bibnamefont {{Takahashi}}},
  \bibinfo {author} {\bibfnamefont {J.~G.}\ \bibnamefont {{Checkelsky}}},
  \bibinfo {author} {\bibfnamefont {N.}~\bibnamefont {{Nagaosa}}}, \bibinfo
  {author} {\bibfnamefont {M.}~\bibnamefont {{Kawasaki}}},\ and\ \bibinfo
  {author} {\bibfnamefont {Y.}~\bibnamefont {{Tokura}}},\ }\bibfield  {title}
  {\bibinfo {title} {{Quantum Hall effect on top and bottom surface states of
  topological insulator (Bi$_{1-x}$Sb$_{x}$)$_{2}$Te$_{3}$ films}},\ }\href
  {https://doi.org/10.1038/ncomms7627} {\bibfield  {journal} {\bibinfo
  {journal} {Nature Communications}\ }\textbf {\bibinfo {volume} {6}},\
  \bibinfo {eid} {6627} (\bibinfo {year} {2015})},\ \Eprint
  {https://arxiv.org/abs/1409.3326} {arXiv:1409.3326 [cond-mat.mtrl-sci]}
  \BibitemShut {NoStop}%
\bibitem [{\citenamefont {Yasuda}\ \emph {et~al.}(2016)\citenamefont {Yasuda},
  \citenamefont {Tsukazaki}, \citenamefont {Yoshimi}, \citenamefont
  {Takahashi}, \citenamefont {Kawasaki},\ and\ \citenamefont
  {Tokura}}]{prl2016_Tokura}%
  \BibitemOpen
  \bibfield  {author} {\bibinfo {author} {\bibfnamefont {K.}~\bibnamefont
  {Yasuda}}, \bibinfo {author} {\bibfnamefont {A.}~\bibnamefont {Tsukazaki}},
  \bibinfo {author} {\bibfnamefont {R.}~\bibnamefont {Yoshimi}}, \bibinfo
  {author} {\bibfnamefont {K.~S.}\ \bibnamefont {Takahashi}}, \bibinfo {author}
  {\bibfnamefont {M.}~\bibnamefont {Kawasaki}},\ and\ \bibinfo {author}
  {\bibfnamefont {Y.}~\bibnamefont {Tokura}},\ }\bibfield  {title} {\bibinfo
  {title} {Large unidirectional magnetoresistance in a magnetic topological
  insulator},\ }\href {https://doi.org/10.1103/PhysRevLett.117.127202}
  {\bibfield  {journal} {\bibinfo  {journal} {Phys. Rev. Lett.}\ }\textbf
  {\bibinfo {volume} {117}},\ \bibinfo {pages} {127202} (\bibinfo {year}
  {2016})}\BibitemShut {NoStop}%
\bibitem [{\citenamefont {Liu}\ \emph {et~al.}(2014)\citenamefont {Liu},
  \citenamefont {Sanderson}, \citenamefont {Cao},\ and\ \citenamefont
  {Zhang}}]{nl_opto_granphene_2014}%
  \BibitemOpen
  \bibfield  {author} {\bibinfo {author} {\bibfnamefont {Z.}~\bibnamefont
  {Liu}}, \bibinfo {author} {\bibfnamefont {M.}~\bibnamefont {Sanderson}},
  \bibinfo {author} {\bibfnamefont {J.~C.}\ \bibnamefont {Cao}},\ and\ \bibinfo
  {author} {\bibfnamefont {C.}~\bibnamefont {Zhang}},\ }\bibfield  {title}
  {\bibinfo {title} {Topologically guaranteed enhancement of nonlinear optical
  conductivity of graphene in the presence of spin-orbit coupling},\ }\href
  {https://doi.org/10.1103/PhysRevB.90.235430} {\bibfield  {journal} {\bibinfo
  {journal} {Phys. Rev. B}\ }\textbf {\bibinfo {volume} {90}},\ \bibinfo
  {pages} {235430} (\bibinfo {year} {2014})}\BibitemShut {NoStop}%
\bibitem [{\citenamefont {Ma}\ \emph {et~al.}(2018)\citenamefont {Ma},
  \citenamefont {Yu}, \citenamefont {Pan},\ and\ \citenamefont
  {Yao}}]{Dashui_graphene_warping_2018}%
  \BibitemOpen
  \bibfield  {author} {\bibinfo {author} {\bibfnamefont {D.-S.}\ \bibnamefont
  {Ma}}, \bibinfo {author} {\bibfnamefont {Z.-M.}\ \bibnamefont {Yu}}, \bibinfo
  {author} {\bibfnamefont {H.}~\bibnamefont {Pan}},\ and\ \bibinfo {author}
  {\bibfnamefont {Y.}~\bibnamefont {Yao}},\ }\bibfield  {title} {\bibinfo
  {title} {Trigonal warping induced unusual spin texture and strong spin
  polarization in graphene with the rashba effect},\ }\href
  {https://doi.org/10.1103/PhysRevB.97.085416} {\bibfield  {journal} {\bibinfo
  {journal} {Phys. Rev. B}\ }\textbf {\bibinfo {volume} {97}},\ \bibinfo
  {pages} {085416} (\bibinfo {year} {2018})}\BibitemShut {NoStop}%
\bibitem [{\citenamefont {Battilomo}\ \emph {et~al.}(2019)\citenamefont
  {Battilomo}, \citenamefont {Scopigno},\ and\ \citenamefont
  {Ortix}}]{BCD_graphene}%
  \BibitemOpen
  \bibfield  {author} {\bibinfo {author} {\bibfnamefont {R.}~\bibnamefont
  {Battilomo}}, \bibinfo {author} {\bibfnamefont {N.}~\bibnamefont
  {Scopigno}},\ and\ \bibinfo {author} {\bibfnamefont {C.}~\bibnamefont
  {Ortix}},\ }\bibfield  {title} {\bibinfo {title} {Berry curvature dipole in
  strained graphene: A fermi surface warping effect},\ }\href
  {https://doi.org/10.1103/PhysRevLett.123.196403} {\bibfield  {journal}
  {\bibinfo  {journal} {Phys. Rev. Lett.}\ }\textbf {\bibinfo {volume} {123}},\
  \bibinfo {pages} {196403} (\bibinfo {year} {2019})}\BibitemShut {NoStop}%
\bibitem [{\citenamefont {Kapri}\ \emph {et~al.}(2021)\citenamefont {Kapri},
  \citenamefont {Dey},\ and\ \citenamefont {Ghosh}}]{BC_role_2021}%
  \BibitemOpen
  \bibfield  {author} {\bibinfo {author} {\bibfnamefont {P.}~\bibnamefont
  {Kapri}}, \bibinfo {author} {\bibfnamefont {B.}~\bibnamefont {Dey}},\ and\
  \bibinfo {author} {\bibfnamefont {T.~K.}\ \bibnamefont {Ghosh}},\ }\bibfield
  {title} {\bibinfo {title} {Role of berry curvature in the generation of spin
  currents in rashba systems},\ }\href
  {https://doi.org/10.1103/PhysRevB.103.165401} {\bibfield  {journal} {\bibinfo
   {journal} {Phys. Rev. B}\ }\textbf {\bibinfo {volume} {103}},\ \bibinfo
  {pages} {165401} (\bibinfo {year} {2021})}\BibitemShut {NoStop}%
\bibitem [{\citenamefont {Ashcroft}\ and\ \citenamefont
  {Mermin}(1976)}]{Ashcroft_my}%
  \BibitemOpen
  \bibfield  {author} {\bibinfo {author} {\bibfnamefont {N.~W.}\ \bibnamefont
  {Ashcroft}}\ and\ \bibinfo {author} {\bibfnamefont {N.~D.}\ \bibnamefont
  {Mermin}},\ }\href@noop {} {\emph {\bibinfo {title} {Solid State Physics}}}\
  (\bibinfo  {publisher} {New York Holt, Rinehart and Winston},\ \bibinfo
  {year} {1976})\BibitemShut {NoStop}%
\bibitem [{\citenamefont {Zeng}\ \emph {et~al.}(2020)\citenamefont {Zeng},
  \citenamefont {Nandy},\ and\ \citenamefont {Tewari}}]{czeng_prr}%
  \BibitemOpen
  \bibfield  {author} {\bibinfo {author} {\bibfnamefont {C.}~\bibnamefont
  {Zeng}}, \bibinfo {author} {\bibfnamefont {S.}~\bibnamefont {Nandy}},\ and\
  \bibinfo {author} {\bibfnamefont {S.}~\bibnamefont {Tewari}},\ }\bibfield
  {title} {\bibinfo {title} {Fundamental relations for anomalous thermoelectric
  transport coefficients in the nonlinear regime},\ }\href
  {https://doi.org/10.1103/PhysRevResearch.2.032066} {\bibfield  {journal}
  {\bibinfo  {journal} {Phys. Rev. Research}\ }\textbf {\bibinfo {volume}
  {2}},\ \bibinfo {pages} {032066} (\bibinfo {year} {2020})}\BibitemShut
  {NoStop}%
\end{thebibliography}%

\end{document}